\documentclass[pra,reprint,groupedaddress]{revtex4-1}

 \usepackage{comment}
 \usepackage{amsmath}
 \usepackage{amsfonts}
 \usepackage{graphicx}
 \usepackage{comment}
 \usepackage{newtxtext,newtxmath}
\usepackage{dcolumn}

 \usepackage{hyperref}
 \hypersetup{
    colorlinks=true,       
    linkcolor=blue,          
    citecolor=blue,        
    filecolor=magenta,      
    urlcolor=blue           
}

\newcommand{\tj}[6]{\begin{pmatrix} #1 & #2 & #3 \\ #4 & #5 & #6 \end{pmatrix}}
\newcommand{\sj}[6]{\begin{Bmatrix} #1 & #2 & #3 \\ #4 & #5 & #6 \end{Bmatrix}}

\newcommand{\oze}{{}^1Z_\text{eff}}
\renewcommand{\vec}[1]{\boldsymbol{\mathbf{#1}}}

\renewcommand{\epsilon}{\varepsilon}

\allowdisplaybreaks

\newcolumntype{d}[1]{D{.}{.}{#1}}

\let\originalleft\left
\let\originalright\right
\renewcommand{\left}{\mathopen{}\mathclose\bgroup\originalleft}
\renewcommand{\right}{\aftergroup\egroup\originalright}

\begin{document}

\frenchspacing

\title{Many-body theory for positronium scattering and pickoff annihilation in noble-gas atoms}
\author{A.~R. Swann}
\email{a.swann@qub.ac.uk}
\author{D.~G. Green}
\email{d.green@qub.ac.uk}
\author{G.~F. Gribakin}
\email{g.gribakin@qub.ac.uk}
\affiliation{
Centre for Theoretical Atomic, Molecular and Optical Physics, School of Mathematics and Physics, Queen's University Belfast, University Road, Belfast BT7 1NN, United Kingdom}
\date{\today}

\begin{abstract}
The many-body-theory approach to positronium-atom interactions developed in [Phys. Rev. Lett. \textbf{120}, 183402 (2018)] is applied to the sequence of noble-gas atoms He--Xe. 
The Dyson equation is solved separately for an electron and positron moving in the field of the atom, with the entire system enclosed in a hard-wall spherical cavity. 
The two-particle Dyson equation is solved to give the energies and wave functions of the Ps eigenstates in the cavity. From these, we determine the scattering phase shifts and cross sections, and values of the pickoff annihilation parameter $\oze$ including short-range electron-positron correlations via vertex enhancement factors. Comparisons are made with available experimental data for elastic and momentum-transfer cross sections and $\oze$. Values of $\oze$ for He and Ne, previously reported in [Phys. Rev. Lett. \textbf{120}, 183402 (2018)], are found to be in near-perfect agreement with experiment, and for Ar, Kr, and Xe within a factor of 1.2.  \end{abstract}

\maketitle

\section{Introduction}
Positronium (Ps) is the bound state of an electron and positron. Being a purely leptonic system, and the simplest ``anti-atom", Ps interactions with matter are of fundamental interest and have applications in many areas \cite{Laricchia12}.
For example, the AEgIS and GBAR experiments at CERN \cite{Kellerbauer08,Debu2012} aim to test whether gravity affects antimatter equivalently to matter, making antihydrogen in Ps collisions with antiprotons, with Ps produced in a mesoporous material \cite{Consolati13}. 
Ps is widely used in condensed-matter physics to determine pore sizes in nanoporous materials and to probe intermolecular voids in polymers \cite{Gidley06}. Moreover, Ps formation in porous materials is used to study its interactions with gases, e.g., Xe \cite{Shibuya13,Shibuya13a}, or the interaction between the Ps atoms themselves \cite{Cassidy07,Cassidy08,Cassidy11,Cassidy12}, with prospects of room-temperature Bose-Einstein condensation and an annihilation gamma laser \cite{Platzman94,Mills07,Cassidy07a}. There are also proposals for using a beam of long-lived Rydberg Ps for measuring the free fall of a matter-antimatter system \cite{Cassidy14} and for detecting positron-atom bound states \cite{Swann16}.

The theoretical description of Ps-atom interactions is challenging due to the composite nature of both scattering objects and a significant cancellation between the short-range Ps-atom repulsion (which results from the exchange interaction between the electrons in the target atom and the electron in Ps) and the long-range van der Waals attraction. Accurate calculations must account for virtual excitation of both objects during the collision.

In this work we carry out calculations of Ps scattering by noble-gas atoms at low energies (i.e., below the Ps ionization potential, 6.8~eV). The authors have previously considered this problem in the frozen-target approximation, where virtual excitations of the target atom are neglected, and also using a model van der Waals potential to approximately account for such excitations~\cite{Swann18}. The calculated scattering cross sections did not agree with the experimental data for Ar and Xe, which indicated that the cross section becomes very small at low Ps energies, suggesting that a Ramsaeur-Townsend minimum may be present~\cite{Brawley15}. 
In a subsequent publication, we 
developed a many-body theory description of Ps-atom interactions by combining the many-body theory
description of electron-atom and positron-atom interactions, while including the important effect of screening
of the electron-positron Coulomb interaction by the atom ~\cite{Green18}. 
As first applications, we computed scattering cross sections and pickoff annihilation rates $\oze$ for Ps collisions with He and Ne~\cite{Green18}. 
The cross section for both targets was found to be an overall rather featureless  curve gently decreasing with higher momenta, with no appearance of a Ramsauer-Townsend minimum. The scattering lengths were found to be positive (1.70~a.u. for He and 1.76~a.u. for Ne), indicating that the Ps-atom interaction is overall repulsive at low energy.
The calculations of $\oze$ were the first to account for short-range electron-positron correlations, which are known to enhance  annihilation rates by a factor of 2--5~\cite{Green15,Green18a}. Including the short-range enhancement gave agreement with experimental measurements of $\oze$ for He and Ne at room temperature~\cite{Charlton85} at the level of 5--10\% accuracy. Here we describe the many-body-theory approach in more detail and extend its application to Ar, Kr and Xe, including a study of the sensitivity of the results on the energies at which the screened Coulomb interaction is calculated.

Unless otherwise stated, atomic units (a.u.) are used throughout, with the symbol $a_0$ denoting the Bohr radius (the atomic unit of length).

\section{Theory}

\subsection{Hard-wall confinement}

The system under consideration is a ground-state Ps atom moving in the field of a closed-shell many-electron atom. We enclose the entire system by an impenetrable spherical wall of radius $R_c$ centered on the target atom. This has the effect of making all Ps states discrete~\cite{Brown17}. This hard-wall cavity is a key feature of our method. Values of $R_c$ are chosen in such a way that the cavity does not affect the atomic ground state and allows for an accurate description of the distortion of Ps as it scatters on the target atom. This enables us to determine Ps-atom scattering phase shifts from the discrete energy eigenvalues~\cite{Swann18}. 

We construct the Ps wavefunction as an expansion of electron and positron states that are solutions of 
 the Dyson equation, which involves the self-energy for the respective particle in the field of the atom. The Ps state satisfies a two-particle Dyson equation (Bethe-Salpeter equation), which we solve to find the expansion coefficients and discrete energy eigenvalues from which the scattering phase shifts can be determined~\cite{Swann18}. We now describe each step in detail. 

\subsection{Dyson equation for electron or positron}

A conventional treatment of an electron or positron interaction with an $N$-electron atom would start from the Schr\"odinger equation for the total wave function for the $N+1$ particles. In many-body theory we instead start from the Dyson equation (see, e.g., Refs.~\cite{fetterwalecka,boylepindzola}):
\begin{equation}\label{eq:el_pos_dyson}
(H_0 + \Sigma_E)\psi_E = E\psi_E.
\end{equation}
Here, $\psi_E$ is the single-particle (quasiparticle) wave function of the incident electron or positron, $E$ is its energy, and $H_0$ is a central-field Hamiltonian that describes the motion of the incident electron or positron in the static field of the atom (including exchange for the electron). The key quantity in Eq.~(\ref{eq:el_pos_dyson}) is $\Sigma_E$, a nonlocal, energy-dependent correlation potential that is equal to the self-energy part of the single-particle Green's function of the electron or positron in the field of the atom~\cite{Bell59}. Due to its nonlocal, dynamical nature, $\Sigma_E$ operates on $\psi_E$ as an integral operator:
\begin{equation}
\Sigma_E \psi_E = \int \Sigma_E(\vec{r},\vec{r}')\psi_E(\vec{r}') \, d^3\vec{r}'.
\end{equation}

The self-energy $\Sigma_E$ is given by an infinite series in powers of the residual electron-electron and/or electron-positron interactions. The use of the Hartree-Fock approximation for the atomic electrons and inclusion of the electrostatic (and exchange) interaction in $H_0$ means that the expansion for $\Sigma_E$ starts with the second-order diagrams, and the diagrams do not contain elements that describe the electrostatic potential of the atom~\cite{Gribakin04}. For electrons, this also implies the absence from $\Sigma_E$ of the contribution of the target exchange potential.

As a result of the spherical symmetry of the problem, Eq.~(\ref{eq:el_pos_dyson}) can be solved separately for each partial wave of the incident electron or positron. The self-energy is expanded in partial waves as
\begin{equation}
\Sigma_E(\vec{r},\vec{r}') = \frac{1}{rr'} \sum_{\lambda=0}^\infty \Sigma_E^{(\lambda)}(r,r')\sum_{\mu=-\lambda}^\lambda Y_{\lambda\mu}^*(\vec{\hat{r}}) Y_{\lambda\mu}(\vec{\hat{r}}'),
\end{equation}
where $Y_{\lambda\mu}$ is a spherical harmonic. Rather than using the coordinate representation $\Sigma_E(\vec{r},\vec{r}')$ of the self-energy, it is usually more convenient to work with its matrix elements in the basis of eigenfunctions of $H_0$, viz.,
\begin{align}
\langle \epsilon' l' m' \vert \Sigma_E \vert \epsilon l m \rangle 
&=
\iint \varphi^*_{\epsilon'l'm'}(\vec{r}') \Sigma_E(\vec{r},\vec{r}') \varphi_{\epsilon l m}(\vec{r}) \, d^3\vec{r} \, d^3\vec{r}' \notag\\
&=
\delta_{ll'}\delta_{mm'} \iint P_{\epsilon'l'}(r') \Sigma_E^{(l)}(r,r') P_{\epsilon l}(r) \, dr \, dr' , \label{eq:self_energy_matrix_coord}
\end{align}
where
\begin{align}
H_0 \varphi_{\epsilon l m}(\vec{r}) &= \epsilon \varphi_{\epsilon l m}(\vec{r}) , \\
\varphi_{\epsilon l m}(\vec{r}) &= \frac{1}{r} P_{\epsilon l}(r) Y_{lm}(\vec{\hat{r}}).
\end{align}
For brevity, we will frequently abbreviate the set of quantum numbers $\epsilon l m$ by the single label $\epsilon$.

\subsection{Calculation of the self-energy}

\begin{figure*}
\includegraphics[scale=.7]{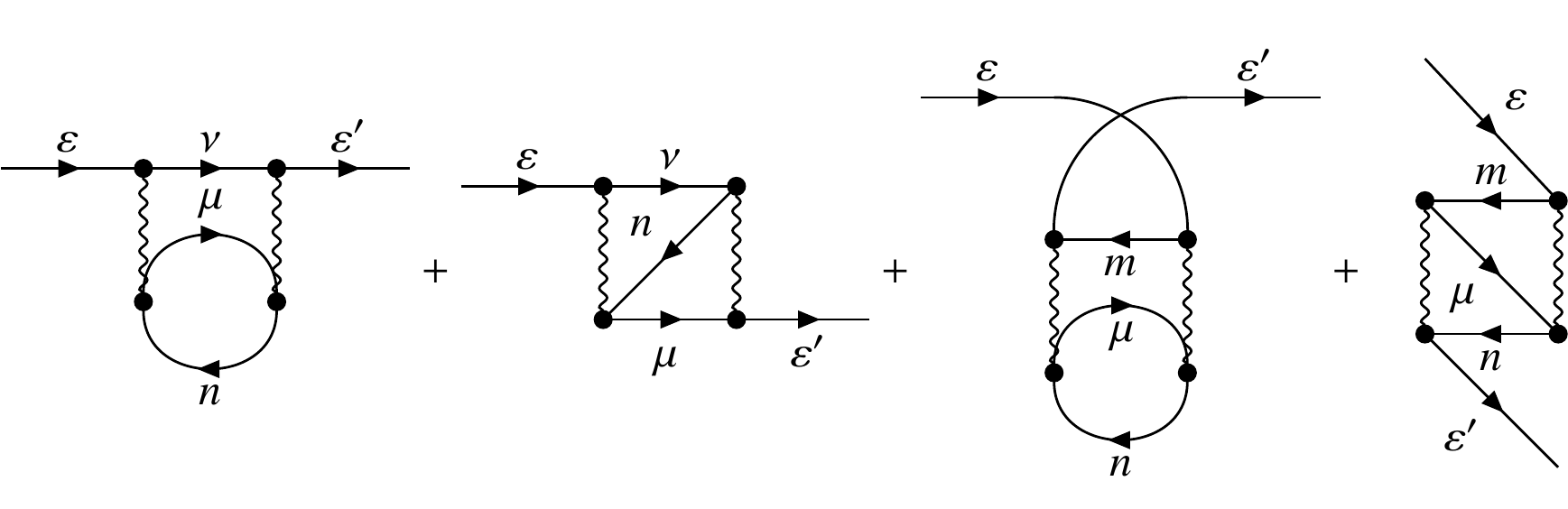}\\
\includegraphics[scale=.7]{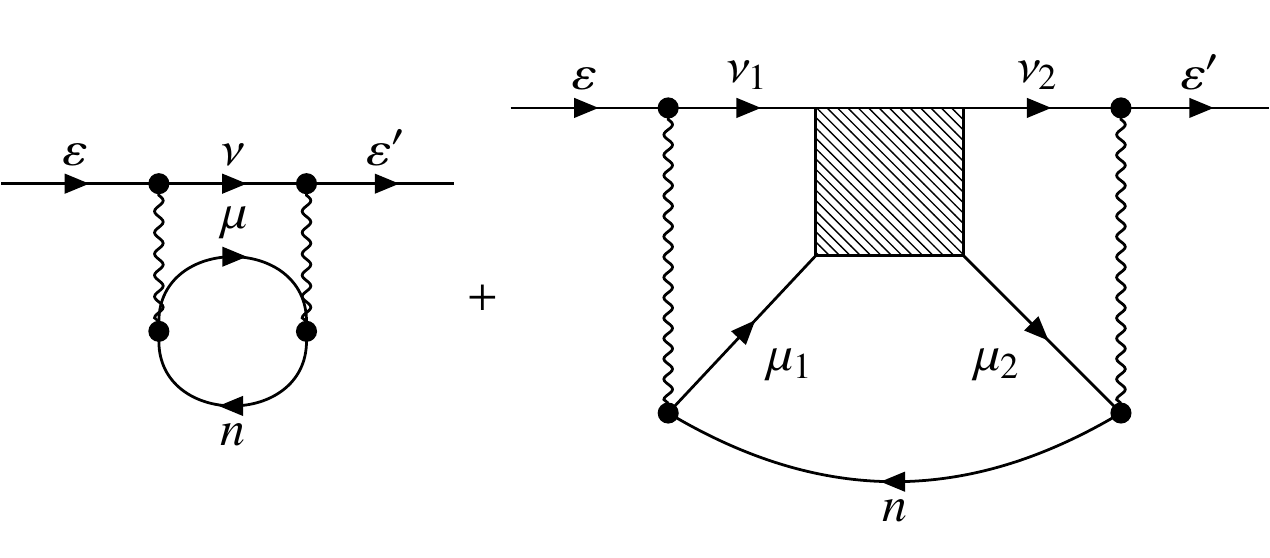}
\caption{\label{fig:el_pos_diagrams}The main contributions to $\langle \epsilon' \vert \Sigma_E \vert \epsilon \rangle $ for the electron (top row) and positron (bottom row). Lines labeled $\epsilon$ or $\epsilon'$ represent the electron or positron  wave function in the static field of the atom. Internal lines labeled $\mu$ or $\nu$ represent either excited electron or positron states, while those labeled $m$ or $n$ represent holes in the atomic ground state. Wavy lines represent electron-electron and electron-positron Coulomb interactions. The hatched block represents the electron-positron ladder-diagram series (see Fig.~\ref{fig:ladder}).}
\end{figure*}

Each contribution to $\langle \epsilon' \vert \Sigma_E \vert \epsilon \rangle $ can be represented by a Goldstone diagram. For the electron case we only include diagrams of the lowest, second order, as shown in Fig.~\ref{fig:el_pos_diagrams} (top row). The second-order diagrams are known to provide an accurate description of electron-atom interactions~\cite{Amusia74,Amusia75,Amusia82,Johnson94,Johnson01}.
The first of these diagrams has the following analytical expression:
\begin{equation}\label{eq:pol_dia_expr}
\langle \epsilon' \vert \Sigma_E \vert \epsilon \rangle =
\sum_{\substack{\mu,\nu> F \\ n\leq F}} \frac{\langle \epsilon' n \vert V \vert \mu \nu \rangle \langle \nu \mu \vert V \vert n \epsilon \rangle}{E+\epsilon_n-\epsilon_\mu-\epsilon_\nu+i\delta} ,
\end{equation}
where $V=\lvert \vec{r}-\vec{r}'\rvert^{-1}$ is the repulsive electron-electron Coulomb interaction, with matrix elements defined as
\begin{equation}\label{eq:coulomb_matrix_element}
\langle \nu \mu \vert V \vert n \epsilon \rangle
=
\iint \frac{\varphi_\nu^*(\vec{r}') \varphi_\mu^*(\vec{r}) \varphi_n(\vec{r}) \varphi_\epsilon(\vec{r})}{\lvert \vec{r}-\vec{r}'\rvert} \, d^3\vec{r} \, d^3\vec{r}';
\end{equation}
$\epsilon_\mu$ is the energy of state $\mu$, etc.; $\delta$ is a positive infinitesimal; and the sums run over all  holes $n$  
and all excited electron states $\mu$, $\nu$,
including integration over the continuum (although due to the presence of the hard spherical wall, the positive-energy ``continuum'' states are discrete). This diagram accounts for the main correlation effect in low-energy electron-atom interactions, namely, polarization of the atom. At large distances, it leads to the well-known local polarization potential,
\begin{equation}
\Sigma_E(\vec{r},\vec{r}') \simeq -\frac{\alpha}{2r^4} \delta(\vec{r}-\vec{r}'),
\end{equation}
where
\begin{equation}
\alpha = \frac{2}{3} \sum_{\substack{\mu>F \\ n\leq F}} \frac{\lvert \langle \mu \vert \vec{r} \vert n \rangle \rvert^2}{\epsilon_\mu - \epsilon_n}
\end{equation}
is the static dipole polarizability of the atom in the Hartree-Fock approximation. The other three diagrams only contribute to $\langle \epsilon' \vert \Sigma_E \vert \epsilon \rangle$ at short range.

\begin{figure*}
\includegraphics[scale=.7]{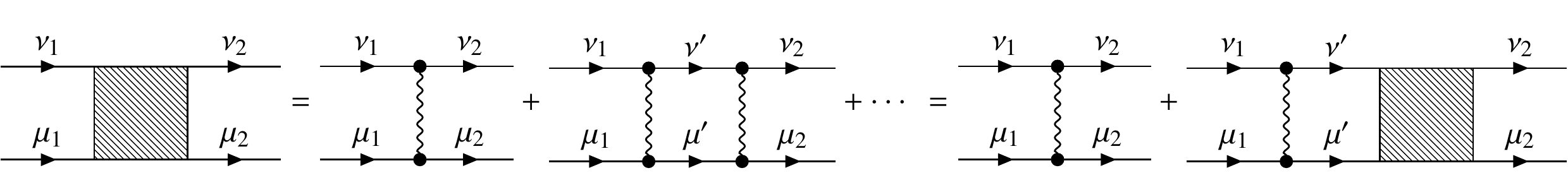}
\caption{\label{fig:ladder}The electron-positron ladder-diagram series, which accounts for virtual Ps formation.}
\end{figure*}

The main contributions to $\langle \epsilon' \vert \Sigma_E \vert \epsilon \rangle$ for the positron are shown in Fig.~\ref{fig:el_pos_diagrams} (bottom row). The first diagram produces a long-range polarization potential, similar to that for the electron. The second diagram describes the important effect of virtual Ps formation~\cite{Dzuba93,Gribakin04,Green14}; the hatched block represents the sum of the infinite electron-positron ladder-diagram series, as shown in Fig.~\ref{fig:ladder}.
This infinite series may be calculated by considering the analytical form of Fig.~\ref{fig:ladder},
\begin{align}
\langle \nu_2 \mu_2 \vert \Gamma_E \vert \mu_1 \nu_1 \rangle
&=
-\langle \nu_2 \mu_2 \vert V \vert \mu_1 \nu_1 \rangle \notag\\
&\quad{}-
\sum_{\mu',\nu'} \frac{\langle \nu_2 \mu_2 \vert \Gamma_E \vert \mu' \nu' \rangle \langle \nu' \mu' \vert V \vert \mu_1 \nu_1 \rangle}{E-\epsilon_{\mu'}-\epsilon_{\nu'}+i\delta}, \label{eq:ladder}
\end{align}
where $-V$ is the attractive electron-positron Coulomb interaction. Equation~(\ref{eq:ladder}) is an integral equation for the ladder matrix elements. Due to the hard spherical wall (which discretizes the electron and positron continua), it becomes a linear matrix equation, which can  be solved to find the ladder matrix elements~\cite{Gribakin04}.

See Appendix~\ref{sec:analytical_expressions} for analytical expressions for each of the diagrams shown in Fig.~\ref{fig:el_pos_diagrams}.

\subsection{Electron and positron quasiparticle wave functions and energies}

In either the electron or positron case, we calculate the self-energy matrix elements $\langle \epsilon' \vert \Sigma_E \vert \epsilon \rangle$ for pairs of Hartree-Fock states $\epsilon l m$ and $\epsilon' l m$ in the hard-wall cavity. Then, we determine the quasiparticle wave functions and energies as follows. We expand the unknown quasiparticle wave function (also called a \textit{Dyson orbital}) with energy $E$ in the basis of Hartree-Fock wave functions,
\begin{equation}\label{eq:dys_orb_exp}
\psi_{Elm}(\vec{r}) = \sum_{\epsilon} C_{\epsilon} \varphi_{\epsilon l m}(\vec{r}).
\end{equation}
The function $\psi_{Elm}(\vec{r})$ can be factorized into radial and angular parts, 
\begin{equation}
\psi_{Elm}(\vec{r}) = \frac1r \mathcal{P}_{El}(r)Y_{lm}(\vec{\hat{r}}),
\end{equation}
so Eq.~(\ref{eq:dys_orb_exp}) is really just an expansion for the radial part:
\begin{equation}
\mathcal{P}_{El}(r) = \sum_{\epsilon} C_{\epsilon} P_{\epsilon l}(r).
\end{equation}
Substituting Eq.~(\ref{eq:dys_orb_exp}) into the Dyson equation (\ref{eq:el_pos_dyson}) and taking matrix elements, we obtain a matrix-eigenvalue equation
\begin{equation}\label{eq:el_pos_eiv_eq}
\vec{H} \vec{C} = E \vec{C},
\end{equation}
where the Hamiltonian matrix $\vec{H}$ has elements
\begin{equation}\label{eq:el_pos_ham_mat}
\langle \epsilon' \vert H \vert \epsilon \rangle = \epsilon \delta_{\epsilon \epsilon'} + \langle \epsilon' \vert \Sigma_E \vert \epsilon \rangle,
\end{equation}
and $\vec{C}$ is the vector of expansion coefficients $C_\epsilon$. Solving Eq.~(\ref{eq:el_pos_eiv_eq}) for each partial wave yields the Dyson-orbital energies and corresponding expansion coefficients for the quasiparticle wave functions. For the electron, the result is a set of negative-energy states corresponding to the atomic orbitals, along with a set of positive-energy ``contimuum'' states. For the positron, all of the states are positive-energy ``continuum'' states.

\subsection{\label{sec:en_dep_self_energy}Energy dependence of the self-energy}

Because of the dynamic nature of the electron- or positron-atom correlation potential, the self-energy is energy dependent: the energy $E$ appears in the energy denominators of the Goldstone diagrams [see, e.g., Eq.~(\ref{eq:pol_dia_expr})]. However, in finding the Dyson orbitals we do not know the value of $E$ until we have solved Eq.~(\ref{eq:el_pos_ham_mat}), which requires the matrix elements $\langle \epsilon' \vert \Sigma_E \vert \epsilon \rangle$ to already have been calculated. This problem could be circumvented by solving the Dyson equation self-consistently with some initial guess for the energy $E$ being used in $\langle \epsilon' \vert \Sigma_E \vert \epsilon \rangle$ (e.g., the corresponding Hartree-Fock energy). However, there are at least two difficulties associated with such an approach:
\begin{enumerate}
\item For any given partial wave of the incident electron or positron, the self-consistent solution of the Dyson equation would have to be carried out separately for each radial state, i.e., when solving Eq.~(\ref{eq:el_pos_ham_mat}) only the energy and expansion coefficients for the state under consideration are accurate; the energies and expansion coefficients that arise for the other states must be ignored.
\item If each of the Dyson orbitals is calculated with its own self-energy matrix $\langle \epsilon' \vert \Sigma_E \vert \epsilon \rangle$ then they will not be mutually orthogonal. Hence they will not be suitable for constructing the two-particle Ps wave function (see Sec.~\ref{sec:2part_Ps_states}).
\end{enumerate}
In light of this, we have chosen to consistently calculate all of the self-energy diagrams at $E=0$. The energy dependence of the self-energy matrix elements $\langle \epsilon' \vert \Sigma_E \vert \epsilon \rangle$ is rather weak in the energy range of interest and provides a  good description of electron and positron interactions with noble-gas atoms~\cite{Green18}. In Sec.~\ref{sec:results_energy_dependence} we will briefly investigate the effect of changing the value of $E$ on the results.

\subsection{\label{sec:2part_Ps_states}Two-particle Ps states}

\begin{figure*}
\includegraphics[width=\textwidth]{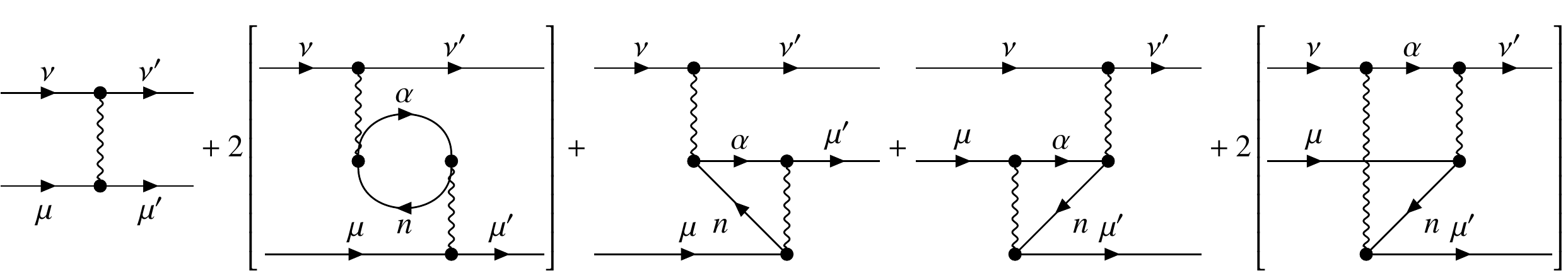}
\caption{\label{fig:screening_diagrams}The main contributions to $\langle \nu' \mu' \vert {-}V+\delta V_E \vert \mu \nu \rangle$. The top line (with end states $\nu$ and $\nu'$) represents the positron (electron), while the bottom line (with end states $\mu$ and $\mu'$) represents the electron. The factors of 2 are to account for mirror-image diagrams that have identical analytical expressions. The first diagram is the bare Coulomb-interaction matrix element $\langle \nu' \mu' \vert {-}V \vert \mu \nu \rangle$. The second diagram is the \textit{direct} screening diagram; the other diagrams are \textit{exchange} screening diagrams.}
\end{figure*}

Having determined the Dyson-orbital energies and wave functions for both the electron and positron cases, one can construct a two-particle Ps wave function. The two-particle Dyson equation (also known as the Bethe-Salpeter equation~\cite{fetterwalecka}) for Ps moving in the field of the target atom is
\begin{equation}\label{eq:bethe_salpeter}
(H_0^e + H_0^p + \Sigma_{E^e}^e + \Sigma_{E^p}^p - V + \delta V_E)\Psi = E\Psi ,
\end{equation}
where $H_0^e$ ($H_0^p$) is the single-particle electron (positron) Hamiltonian [which includes the electron (positron) kinetic energy and the Hartree-Fock potential of the atom], $\Sigma_{E^e}^e$ ($\Sigma_{E^p}^p$) is the self-energy of the electron (positron), $-V$ is the attractive Coulomb interaction between the electron and positron in Ps, and $\Psi$ is the two-particle Ps wave function with energy $E$ \footnote{There is a similarity between our approach and the combination of MBT with the configuration-interaction method for open-shell atoms \cite{Dzuba:1996}}. 
The operator $\delta V_E$ is the screening correction to the electron-positron Coulomb interaction. The diagrams for the matrix elements $\langle \nu' \mu' \vert {-}V+\delta V_E \vert \mu \nu \rangle$ are shown in Fig.~\ref{fig:screening_diagrams}; again, we calculate diagrams up to second order.
The \textit{direct} screening diagram (the second diagram in Fig.~\ref{fig:screening_diagrams}) is essential for canceling the long range $r^{-4}$ behavior of the single-particle electron and positron polarization diagrams and making the long-range Ps-atom interaction of the required $R^{-6}$ van der Waals form, where $R$ is the distance between the nucleus of the target atom and the Ps center of mass. The \textit{exchange} screening diagrams (the third, fourth, and fifth diagrams in Fig.~\ref{fig:screening_diagrams}) are more  expensive to calculate than the direct diagram, but they partially cancel each other are almost negligible in comparison  (see Sec.~\ref{sec:results_scattering} and Fig.~\ref{fig:Ar_different_approximations}); consequently, we will not include the exchange screening diagrams in our calculations.

From this point forward, we shall, without exception, use the label $\mu$ to refer to electron states in the field of the atom, and we shall use the label $\nu$ to refer to positron states in the field of the atom.
Since the   electron and positron Dyson orbitals $\psi_\mu(\vec{r}_e)$ and $\psi_\nu(\vec{r}_p)$ are eigenstates of the single-particle Dyson equations (\ref{eq:el_pos_dyson}), they form mutually orthonormal sets. From them, a two-particle Ps wave function with fixed total angular momentum $J$ and parity $\Pi$ may be constructed as 
\begin{equation}\label{eq:Ps_wfn}
\Psi_{J\Pi}(\vec{r}_e,\vec{r}_p) = \sum_{\mu,\nu} C_{\mu\nu} \psi_\mu(\vec{r}_e) \psi_\nu(\vec{r}_p) ,
\end{equation}
where the $C_{\mu\nu}$ are expansion coefficients.
Note that the electron in Ps should be orthogonal to the atomic electrons; hence, the negative-energy electron states corresponding to the atomic orbitals are excluded from Eq.~(\ref{eq:Ps_wfn}).
The energy eigenvalues $E$ and expansion coefficients $C_{\mu\nu}$ are found by solving the matrix-eigenvalue problem for the Hamiltonian matrix
\begin{equation}\label{eq:Ps_ham_mat}
\langle \nu' \mu' \vert H \vert \mu \nu \rangle
=
(\epsilon_\mu + \epsilon_\nu) \delta_{\mu \mu'} \delta_{\nu \nu'} 
+ 
\langle \nu' \mu' \vert {-}V + \delta V_E \vert \mu \nu \rangle ,
\end{equation}
where $\epsilon_\mu$ is the energy of state $\psi_\mu$, etc.
We consider $J^\Pi=0^+$, $1^-$, and $2^+$ to investigate Ps $S$-, $P$-, and $D$-wave scattering, respectively. The Ps energy eigenvalues in the cavity can be used to determine the scattering phase shifts, and hence the scattering cross section; see Ref.~\cite{Swann18} for details.

\subsection{Pickoff annihilation}

The Ps pickoff annihilation rate $\lambda$ in a gas is parameterized as 
\begin{equation}
\lambda = 4\pi r_0^2 c n \oze ,
\end{equation}
where $r_0$ is the classical electron radius, $c$ is the speed of light, and $n$ is the number density of the gas. The parameter $\oze$ (or, more specifically, $4\oze$) represents the effective number of atomic electrons available for pickoff annihilation~\cite{Fraser66}. 
In the zeroth-order, independent-particle approximation, $\oze$ is given by
\begin{equation}\label{eq:1zeff_def}
\oze = \frac14 \iiint \rho(\vec{r}) \lvert \Psi(\vec{r}_e,\vec{r}_p) \rvert^2 \delta(\vec{r}-\vec{r}_p) \, d^3\vec{r} \, d^3\vec{r}_e \, d^3\vec{r}_p ,
\end{equation}
where $\rho(\vec{r})=\sum_{n\leq F} \lvert \psi_n(\vec{r})\rvert^2$ is the density of the atomic electrons, and $\Psi(\vec{r}_e,\vec{r}_p)$ is the Ps wave function, with the Ps center-of-mass motion normalized to a plane wave at large distances. The factor of $\frac14$ is to account for the fact that only those electrons that form a relative singlet state with the positron contribute to pickoff annihilation (assuming that all annihilation events are the dominant $2\gamma$ decays).

Our interest is in values of $\oze$ at small (thermal) Ps energies, where only the $S$ wave contributes. Higher contributions are suppressed as $K^{2L}$, where $K$ is the Ps momentum and $L$ is the orbital angular momentum of the Ps center-of-mass motion. We therefore use $\Psi=\Psi_{0^+}$ (with appropriate normalization) in Eq.~(\ref{eq:1zeff_def}).

To account for the short-range-correlation corrections to $\oze$ (neglected in the independent-particle approximation), we augment Eq.~(\ref{eq:1zeff_def}) with \textit{vertex enhancement factors} $\gamma_{nl}$, which are specific to the electron orbital $n$ and positron partial wave $l$, and were calculated for positron annihilation in noble-gas atoms in Refs.~\cite{Green15,Green18a}. Explicitly, substituting Eq.~(\ref{eq:Ps_wfn}) into Eq.~(\ref{eq:1zeff_def}), using orthonormality of the electron wave functions, and introducing the enhancement factors yields
\begin{equation}\label{eq:enhanced_oze}
\oze = \frac14 \sum_{n,\mu,\nu,\nu'} \gamma_{nl} C_{\mu\nu} C_{\mu\nu'}^*
\int \lvert \psi_n(\vec{r}) \rvert^2 \psi_\nu(\vec{r}) \psi_{\nu'}^*(\vec{r}) \, d^3\vec{r} ,
\end{equation}
where the positron  states $\psi_\nu$ and $\psi_{\nu'}$ both have angular momentum $l$ (see Appendix~\ref{sec:analytical_expressions} for details of how $\oze$ is computed in practice). The enhancement factors are largest for the valence electrons but may still be significant for the core electrons~\cite{Bonderup79}. 
Also, the values of $\gamma_{nl}$ are dependent on the energy of the incident positron, though this dependence is not very strong (see Fig.~13 in Ref.~\cite{Gribakin04}). Here we use values of $\gamma_{nl}$ at zero energy of the positron.
Note that the $\gamma_{nl}$ can be calculated using either Hartree-Fock or Dyson states for the positron. Here we use the Hartree-Fock values, because when the positron is ``packaged'' within Ps, it is shielded from the target atom by its partner electron. The two particles together cannot polarize the atom as much as a lone positron could, and virtual Ps contributes little because the positron already has a partner electron.
Table~\ref{tab:enhancement_factors} shows these values of $\gamma_{nl}$ (taken from Ref.~\cite{Green18a}).
\begin{table}
\caption{\label{tab:enhancement_factors}Enhancement factors $\gamma_{nl}$ for atomic orbital $n$ and positron partial wave $l$~\cite{Green15,Green18a}.}
\begin{ruledtabular}
\begin{tabular}{lcccc}
Atom & $n$ & $l=0$ & $l=1$ & $l=2$ \\
\hline
He & $1s$ & 2.99 & 4.04 & 5.26 \\
Ne & $1s$ & 1.18 & 1.21 & 1.22 \\
&$2s$ & 1.87 & 2.03 & 2.30 \\
&$2p$ & 2.78 & 3.46 & 4.70 \\ 
Ar & $2s$ & 1.35 & 1.38 & 1.41 \\
&$2p$ & 1.43 & 1.47 & 1.51 \\
&$3s$ & 2.53 & 2.70 & 3.00 \\
&$3p$ & 5.19 & 6.22 & 8.17 \\
Kr & $3s$ & 1.34 & 1.36 & 1.39 \\
& $3p$ & 1.41 & 1.43 & 1.47 \\
& $3d$ & 1.67 & 1.72 & 1.71 \\
& $4s$ & 2.77 & 2.96 & 3.25 \\
& $4p$ & 6.63 & 7.89 & 10.33 \\
Xe & $4s$ & 1.42 & 1.44 & 1.46 \\
& $4p$ & 1.52 & 1.47 & 1.57 \\
& $4d$ & 1.96 & 2.00 & 1.24 \\
& $5s$ & 3.36 & 3.59 & 3.95 \\
& $5p$ & 9.26 & 10.91 & 14.08 
\end{tabular}
\end{ruledtabular}
\end{table}
We leave contributions to $\oze$ for core orbitals and/or for $l\geq 3$ unscaled, as we find that they generally contribute ${\lesssim}1\%$ to $\oze$.

\section{\label{sec:numerical_implementation}Numerical implementation}

\subsection{\textit{B}-spline bases}

First, a standard Hartree-Fock program~\cite{Amusia97} is used to compute the electrostatic potential of the ground-state target atom. Then, a $B$-spline basis is used to calculate the single-particle electron and positron Hartree-Fock wave functions  $\varphi_{\epsilon l m}$ (see, e.g., Ref.~\cite{Gribakin04}). We do this for both the electron and positron using two different sets of $B$ splines: the first set contains 40 splines of order 6, defined in a cavity of radius $R_c=30$~a.u., using an exponential knot sequence (see Eq.~(31) in Ref.~\cite{Gribakin04}); the second set contains 60 splines of order 9, defined in a cavity of radius $R_c=10$, 12, 14, or 16~a.u., using a quadratic-linear knot sequence (see Eq.~(28) in Ref.~\cite{Swann18}). The reason for using two distinct sets of $B$ splines is as follows. Calculation of the Goldstone diagrams appearing in the expansions of the electron and positron self-energies requires summation over a complete set of intermediate states. The exponential knot sequence is well-suited to this, since it provides rapid saturation of the electron and positron continua (see Fig.~6 in Ref.~\cite{Gribakin04}). However, we ultimately want to use the electron and positron states to construct a two-particle Ps wave function; for this we need the single-particle states to accurately represent ``physical'' states in the cavity, for which the quadratic-linear knot sequence is well suited~\cite{Swann18}. The two sets of Hartree-Fock states for both the electron and positron are used to find the Dyson orbitals in the following manner:
\begin{enumerate}
\item The Hartree-Fock wave functions $\varphi_{\epsilon l m}$ of the electron or positron, calculated using the exponential knot sequence, are used to compute the self-energy matrix elements $\langle \epsilon' \vert \Sigma_E \vert \epsilon \rangle$.
\item These matrix elements are used to find the coordinate form of the self-energy using the completeness relation, viz.,
\begin{equation}
\Sigma_E^{(l)}(r,r') = \sum_{\epsilon,\epsilon'} P_{\epsilon' l}(r') \langle \epsilon' \vert \Sigma_E \vert \epsilon \rangle P_{\epsilon l}(r) .
\end{equation}
Here, $P_{\epsilon' l}(r')$ and $P_{\epsilon l}(r)$ are the same Hartree-Fock functions used in step 1, but \textit{evaluated at points $r$ and $r'$ on the quadratic-linear knot sequence}. [See Fig.~2 of \cite{Green18} for illustrative plots of $\Sigma^{(l)}(r,r')$ for $l=0,1,2$ for Ne].
\item The coordinate form of $\Sigma_E$ is used to calculate its matrix elements via Eq.~(\ref{eq:self_energy_matrix_coord}), where now $P_{\epsilon' l}(r')$ and $P_{\epsilon l}(r)$ are the Hartree-Fock functions calculated using the \emph{quadratic-linear} knot sequence. (Steps 1--3 are similarly used later for finding the matrix elements of $\delta V_E$.)
\item The Hamiltonian matrix is diagonalized. Its elements are given by Eq.~(\ref{eq:el_pos_ham_mat}), where the matrix elements $\langle \epsilon' \vert \Sigma_E \vert \epsilon \rangle$ are those calculated in step 3. The eigenenergies are the Dyson-orbital energies of the electron or positron in the cavity, and the eigenvectors provide the expansion coefficients for the quasiparticle wave functions in terms of the Hartree-Fock functions on the quadratic-linear knot sequence.
\end{enumerate} 
This method enables the exponential-sequence Hartree-Fock wave functions to be used in the sums over intermediate states of diagrams, thus ensuring completeness, while the Dyson-orbital energies and wave functions are ultimately calculated using the quadratic-linear sequence, meaning that they can be used to accurately calculate Ps states in the cavity and determine Ps-atom scattering phase shifts.

Note that when we construct the two-particle Ps wave function from the single-particle electron and positron states [Eq.~(\ref{eq:Ps_wfn})], we only use the Dyson-orbital states for the first few partial waves, specifically, $l=0$--3. To solve the Dyson equation for all angular momenta  of the incident electron or positron included in the expansion of the Ps wave function (up to $l=20$; see Sec.~\ref{sec:conv_Ps_wfn}) is computationally expensive and unnecessary. For higher $l$, the centrifugal barrier prevents the electron or positron from approaching the target atom closely, and the effect of correlations (i.e., $\Sigma_E$) is small. Hence, we  just use the Hartree-Fock states for $l\geq 4$.

\subsection{Convergence with respect to number of intermediate states included in diagrams}

The use of $B$ splines means that rapid convergence is achieved in the sums over intermediate states in the Goldstone diagrams with respect to the number of radial states included for a particular angular momentum. However, convergence with respect to the number of partial waves included is slower. The increment to the electron or positron eigenenergies upon increasing the maximum orbital angular momentum from $l-1$ to $l$  behaves as $(l+\frac12)^{-4}$. Therefore, if we include only partial waves up to $l=l_\text{max}$, the energy eigenvalues approach the ultimate $l_\text{max}\to\infty$ values as follows~\cite{Schwartz62,Schwartz63,Kutzelnigg92,Gribakin02}:
\begin{equation}\label{eq:en_extrap}
E(l_\text{max}) = E + \frac{A}{(l_\text{max}+\frac12)^3}.
\end{equation}

In practice, we calculate the self-energy diagrams for $l_\text{max}=7$, 8, 9, and 10, diagonalize the Hamiltonian matrix for each value of $l_\text{max}$, and extrapolate the resulting eigenenergies using the values for $l_\text{max}=9$ and 10 to find the parameters $E$ and $A$ in Eq.~(\ref{eq:en_extrap}). We use 32 radial states for each angular momentum in all calculations of second-order diagrams. 
%
%
We find that extrapolation typically changes the $l_\text{max}=10$ value of the energy by less than $0.1\%$.

When constructing the two-particle Hamiltonian matrix for Ps scattering, we use these extrapolated energies to compute the diagonal elements. The quasiparticle wave functions used in the expansion for the Ps wave function are those obtained for $l_\text{max}=10$.

\subsection{\label{sec:conv_Ps_wfn}Convergence with respect to number of electron and positron states included in Ps wave function}

In Eq.~(\ref{eq:Ps_wfn}), the sums over the electron and positron states should, in theory, run over all orbital angular momenta and radial quantum numbers (up to infinity), but in practice we use finite maximum values $l_\text{max}$ and $n_\text{max}$, respectively \footnote{This $l_\text{max}$ is not related to the $l_\text{max}$ used in summations over intermediate states in the self-energy diagrams.}. The resulting dimension $\mathcal{N}$ of the Hamiltonian matrix (\ref{eq:Ps_ham_mat}) is 
\begin{equation}
\mathcal{N}
= 
\begin{cases}
n_\text{max}^2 (l_\text{max}+1) & \text{for} \quad J^\Pi=0^+, \\
2 n_\text{max}^2 l_\text{max} & \text{for} \quad J^\Pi=1^-, \\
n_\text{max}^2 (3 l_\text{max}-2) & \text{for} \quad J^\Pi=2^+.
\end{cases}
\end{equation}
To keep the size of the calculations manageable, we used $l_\text{max}=n_\text{max}=20$ for $J^\Pi=0^+$, $l_\text{max}=n_\text{max}=18$ for $J^\Pi=1^-$, and $l_\text{max}=n_\text{max}=16$ for $J^\Pi=2^+$. The Ps eigenenergies are then extrapolated to the limits $l_\text{max}\to\infty$ and $n_\text{max}\to\infty$ as explained in Ref.~\cite{Brown17}.

The pickoff annihilation parameter $\oze$ is calculated using Eq.~(\ref{eq:enhanced_oze}) for the lowest-energy state in the cavity. We do this for each cavity radius, $R_c=10$, 12, 14, 16~a.u., giving values of $\oze$ for four values of the Ps center-of-mass momentum $K$. These values are extrapolated in $l_\text{max}$ according to
\begin{equation}
\oze(l_\text{max},n_\text{max}) = \oze(\infty,n_\text{max}) + \frac{A}{(l_\text{max}+\frac12)^2},
\end{equation}
and subsequently in $n_\text{max}$ according to
\begin{equation}
\oze(\infty,n_\text{max}) = \oze + \alpha n_\text{max}^{-\beta},
\end{equation}
where we typically find $\beta\approx 4$.

\subsection{Normalization of Ps wave function in calculation of $\oze$}

In Eq.~(\ref{eq:enhanced_oze}), the Ps wave function needs to be such that the center-of-mass motion is normalized to a plane wave at large distances; however, in our calculations in the cavity, it is normalized as
\begin{equation}
\iint \lvert \Psi_{0+}(\vec{r}_e,\vec{r}_p) \rvert^2 \, d^3\vec{r}_e \, d^3\vec{r}_p = 1.
\end{equation}
Assuming that the Ps wave function can be factorized into internal and center-of-mass parts (see Eq.~(11) in Ref.~\cite{Swann18}), which are themselves normalized to unity in the cavity, the normalization of the center-of-mass motion can be corrected as follows:
\begin{enumerate}
\item The center-of-mass density $\rho_\text{cm}(\vec{r})$ in the cavity is given by
\begin{equation}\label{eq:rhocm}
\rho_\text{cm}(\vec{r}) = \iint \lvert \Psi_{0+}(\vec{r}_e,\vec{r}_p) \rvert^2 \delta \left( \frac{\vec{r}_e+\vec{r}_p}{2}-\vec{r}\right) \, d^3\vec{r}_e \, d^3\vec{r}_p.
\end{equation}
The choice $J=0$ means that $\rho_\text{cm}(\vec{r})$ is spherically symmetric, i.e., it only depends on the distance $r$ of the Ps center of mass from the center of the cavity. We calculate the value of $\rho_\text{cm}$ on a grid from 0 to $R_c$, using integer and half-integer values of $r$. See Appendix~\ref{sec:analytical_expressions} for details of how $\rho_\text{cm}(\vec{r})$ is computed in practice.
\item The value of $\rho_\text{cm}$ for each $r$ is extrapolated to the limit $l_\text{max}\to\infty$, according to 
\begin{equation}
\rho_\text{cm}(l_\text{max}) = \rho_\text{cm} + \frac{A}{(l_\text{max}+\frac12)^3}.
\end{equation}
This extrapolation typically changes the value of $\rho_\text{cm}$ by 1--10\%, depending on the radius $R_c$ of the cavity. In principle one can also extrapolate in $n_\text{max}$; however, it was found that doing so changes the value of $\rho_\text{cm}$ by less than $0.5\%$, so it can be neglected.
\item In the asymptotic region of the cavity, i.e., the region where the Ps center of mass is not too close to either the target atom or the cavity wall \footnote{The asymptotic region is roughly defined by $r_\text{at}\ll r <R_c-\rho_{1s}$, where $r_\text{at}$ is the approximate radius of the target atom, and $\rho_\text{1s}$ is the collisional radius of Ps with respect to the cavity wall~\cite{Brown17}.}, we expect $\rho_\text{cm}(\vec{r})$ to have the form corresponding to free motion:
\begin{equation}\label{eq:rhocm_analytical}
\rho_\text{cm}(\vec{r}) = B^2 \frac{\sin^2(Kr+\delta_0)}{(Kr)^2},
\end{equation}
where $B$ is a normalization constant, $K$ is the Ps center-of-mass momentum, and $\delta_0$ is the $S$-wave scattering phase shift. The value of $B$ is determined by performing a least-squares fit of Eq.~(\ref{eq:rhocm_analytical}) to the calculated values of $\rho_\text{cm}$, with $B$ as a parameter. The phase shift $\delta_0$ can be taken from the scattering calculation or allowed to be a second parameter of the fit. We generally find that allowing $\delta_0$ to be a variable parameter gives a slightly better fit to the calculated values of $\rho_\text{cm}$, so we  allow this in all calculations \footnote{One can alternatively estimate the value of $B^2$ analytically by modeling the target atom as a hard sphere of radius $r_\text{at}$. Assuming $\rho_\text{cm}(\vec{r})=B^2 \sin^2(Kr+\delta_0)/(Kr)^2$ for $r_\text{at}<r<R_c$, and using the fact that $\rho_\text{cm}$ is normalized to unity in the cavity, we obtain $(4\pi B^2/K^2) \int_{r_\text{at}}^{R_c-\rho_{1s}} \sin^2(Kr+\delta_0) \, dr=1$, where $\rho_{1s}$ is the collisional radius of ground-state Ps. Since $r_\text{at}<r<R_c$ corresponds to one half-period of $\sin(Kr+\delta_0)$, i.e., $K(R_c - \rho_{1s}-r_\text{at})=\pi$, we have $\int_{r_\text{at}}^{R_c-\rho_{1s}} \sin^2(Kr+\delta_0) \, dr=\pi /2K$, and therefore $B^2=K^3/2\pi^2$.}. Figure~\ref{fig:cm_density} shows the center-of-mass density for the lowest-energy Ps eigenstate in  collisions with Ar, in a cavity of radius $R_c=10$~a.u., in the frozen-target approximation [where we excluded $\Sigma_E$ for both the electron and positron and $\delta V_E$ from Eq.~(\ref{eq:bethe_salpeter})].
\begin{figure}
\centering
\includegraphics{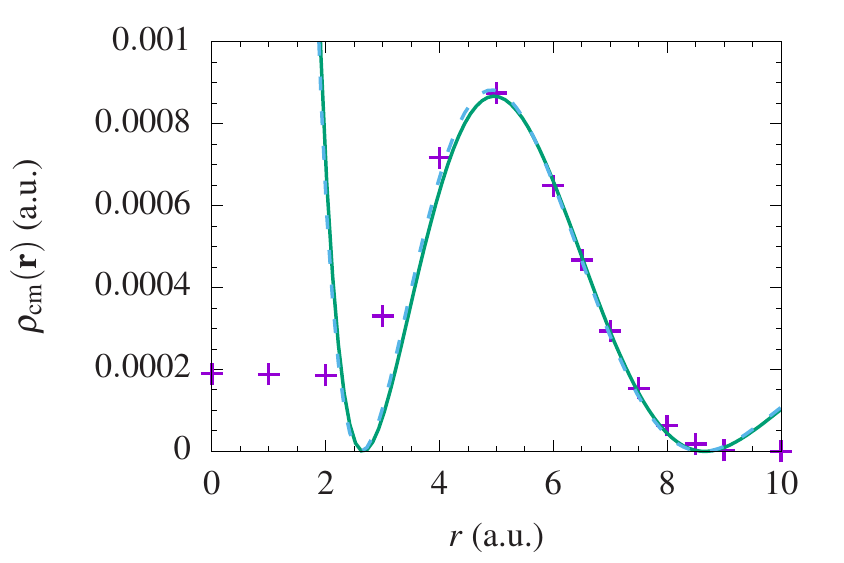}
\caption{\label{fig:cm_density}Center-of-mass density $\rho_\text{cm}(\vec{r})$ for Ps collisions with Ar in a cavity of radius $R_c=10$~a.u., in the frozen-target approximation. Purple plusses, calculated values of $\rho_\text{cm}(\vec{r})$; solid green line, fit using Eq.~(\ref{eq:rhocm_analytical}) with $\delta_0$ as a parameter ($\delta_0=-1.396$); dashed blue line, fit using Eq.~(\ref{eq:rhocm_analytical}) with $\delta_0$ fixed using the boundary condition at the cavity wall ($\delta_0=-1.374$)~\cite{Swann18}. The fits were made using the calculated values of $\rho_\text{cm}(\vec{r})$ for $5\leq r\leq 8.5$~a.u.}
\end{figure}
\item The four values of $\oze$ (one for each of $R_c=10$, 12, 14, 16~a.u.) found directly from Eq.~(\ref{eq:enhanced_oze}) are divided by $B^2$ to obtain correctly normalized values.
\end{enumerate}
Once the four values of $\oze$ have been calculated and correctly normalized, the general dependence of $\oze$ on the Ps momentum $K$ is determined using the effective-range-theory fit
\begin{equation}\label{eq:quadratic_oze}
\oze(K) = \oze(0) + \oze' K^2,
\end{equation}
where $\oze' $ is a parameter to be determined~\cite{Mitroy02b}.

\section{Results}

We have previously reported the  scattering cross sections and values of $\oze$ for He and Ne in Ref.~\cite{Green18}; however, we include them again here for ease of comparison with the current results for Ar, Kr and Xe.

\subsection{\label{sec:results_scattering}Scattering}

\begin{figure*}
\centering
\includegraphics{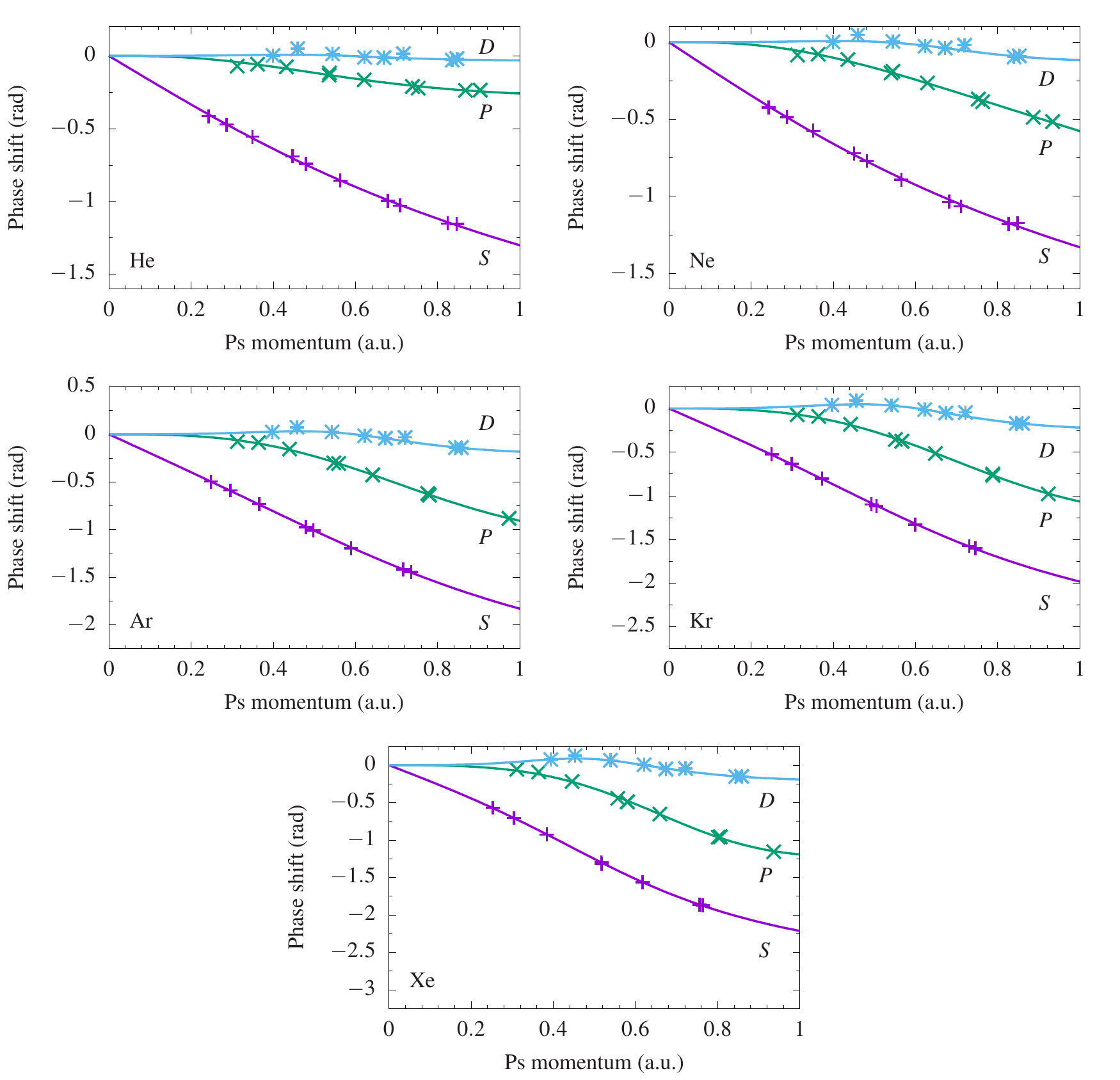}
\caption{\label{fig:phase_shifts_lp00-03hf}$S$-, $P$-, and $D$-wave scattering phase shifts obtained for Ps scattering on He, Ne, Ar, Kr, and Xe. Purple plusses, calculated $\delta_0$; green crosses, calculated $\delta_1$; blue asterisks, calculated $\delta_2$; solid purple lines, effective-range fits  for $\delta_0$; solid green lines, effective-range fits  for $\delta_1$; solid blue lines, effective-range fits  for $\delta_2$ (see Appendix~\ref{sec:ert_fits}).}
\end{figure*}

Figure~\ref{fig:phase_shifts_lp00-03hf} shows the $S$-, $P$-, and $D$-wave scattering phase shifts $\delta_L$ obtained for Ps collisions with He, Ne, Ar, Kr, and Xe. 
The phase shifts are calculated at discrete values of the Ps center-of-mass momentum $K$, and effective-range theory is used to determine the general dependence of $\delta_L$ on $K$; see Appendix~\ref{sec:ert_fits} for details.
In spite of the large differences in the sizes and polarizabilities of the atoms, there is a great degree of similarity between the phase shifts for all of them. This is a result of cancellation of the effect of increasing repulsion in the He to Xe sequence (due to increasing atomic sizes) and increasing strength of the correlation (van der Waals--type) attraction between the Ps and the atom.
Note that the $S$-wave phase shift is negative at low $K$ for all five target atoms, i.e., the scattering length is positive for all of them. Therefore, we can immediately deduce that a Ramsauer-Townsend minimum will not appear in any of our cross sections, in contrast to the experimental prediction~\cite{Brawley15}.

Table~\ref{tab:scattering_lengths} shows the scattering lengths and zero-energy cross sections, along data from a number of previous calculations: the  frozen-target calculations and model van der Waals calculations of Swann and Gribakin~\cite{Swann18}; the coupled-channel $R$-matrix calculation for He (which included 9 Ps states and 9 He states in the channel space) of Walters \textit{et al.}~\cite{Walters04}; fixed-core stochastic-variational calculations for Ne, Ar, Kr, and Xe (which included model polarization potentials for the electron and positron, along with a two-body polarization potential to ensure the correct long-range van der Waals behavior) of Mitroy and coworkers~\cite{Mitroy01,Mitroy03}; and the calculations of Fabrikant and coworkers where the electron- and positron-atom interactions were modeled using pseudopotentials, along with a model long-range van der Waals potential~\cite{Fabrikant14,Gribakin16}.
For Ps scattering on He, we obtain a scattering length of $1.70$~a.u., which is close to our previous  calculation~\cite{Swann18} that mimicked  distortion of the target atom using a model van der Waals potential (with $R_0=3.0$~a.u.; see Eq.~(24) in Ref.~\cite{Swann18}) and the 9-Ps--9-He--state $R$-matrix calculation of Walters \textit{et al.}~\cite{Walters04} at the level of 6\%. This corresponds to a zero-energy cross section of $11.6\pi a_0^2$. 
\begin{table}
\centering
\caption{\label{tab:scattering_lengths}Scattering lengths $A$ (in units of $a_0$) and zero-energy cross sections $\sigma(0)$ (in units of $\pi a_0^2$) for Ps scattering on He, Ne, Ar, Kr, and Xe.}
\begin{ruledtabular}
\begin{tabular}{lcc}
Method & $A$ & $\sigma(0)$ \\
\hline\\[-1.2ex]
\multicolumn{3}{c}{Ps-He calculations} \\
Present, many-body theory & 1.70 & 11.6 \\
Frozen target~\cite{Swann18} & 1.86 & 13.8 \\
van der Waals, $R_0=2.5$~a.u.~\cite{Swann18} & 1.52 & 9.2 \\
van der Waals, $R_0=3.0$~a.u.~\cite{Swann18} & 1.61 & 10.4 \\
$R$ matrix, 9 Ps states, 9 He states~\cite{Walters04} & 1.6 & 9.9 \\[5pt]
\multicolumn{3}{c}{Ps-Ne calculations} \\
 Present, many-body theory & 1.76 & 12.4 \\
Frozen target~\cite{Swann18} & 2.02 & 16.4 \\
van der Waals, $R_0=2.5$~a.u.~\cite{Swann18} & 1.46 & 8.5 \\
van der Waals, $R_0=3.0$~a.u.~\cite{Swann18} & 1.66 & 11.0 \\
 Stochastic variational, van der Waals~\cite{Mitroy01} & 1.55 & 9.6 \\[5pt]
 \multicolumn{3}{c}{Ps-Ar calculations} \\
   Present, many-body theory & 1.98 & 15.6 \\
Frozen target~\cite{Swann18}& 2.81 & 31.6 \\
   van der Waals, $R_0=2.5$~a.u.~\cite{Swann18} & 1.43 & 8.2 \\
van der Waals, $R_0=3.0$~a.u.~\cite{Swann18} & 2.16 & 18.7 \\
 Stochastic variational, van der Waals~\cite{Mitroy01} & 1.79 & 12.8 \\
 Pseudopotential, van der Waals~\cite{Fabrikant14} & 2.14 & 18.3 \\[5pt]
  \multicolumn{3}{c}{Ps-Kr calculations} \\
    Present, many-body theory & 2.06 & 17.0 \\
Frozen target~\cite{Swann18}& 3.11 & 38.7 \\
van der Waals, $R_0=3.0$~a.u.~\cite{Swann18} & 2.26 & 20.4 \\
van der Waals, $R_0=3.5$~a.u.~\cite{Swann18} & 2.56 & 26.2 \\
 Stochastic variational, van der Waals~\cite{Mitroy03} & 1.98 & 15.6 \\
 Pseudopotential, van der Waals~\cite{Fabrikant14} & 2.35 & 22.1 \\[5pt]
   \multicolumn{3}{c}{Ps-Xe calculations} \\
   Present, many-body theory & 2.12 & 18.1 \\  
Frozen target~\cite{Swann18} & 3.65 & 53.3 \\
     van der Waals, $R_0=3.0$~a.u.~\cite{Swann18} & 2.63 & 27.7 \\
  van der Waals, $R_0=3.5$~a.u.~\cite{Swann18} & 2.88 & 33.2 \\
 Stochastic variational, van der Waals~\cite{Mitroy03} & 2.29 & 20.9 \\
 Pseudopotential, van der Waals~\cite{Gribakin16} & 2.45 & 24.0 
\end{tabular}
\end{ruledtabular}
\end{table}
Note that the scattering length has reduced from its frozen-target value~\cite{Swann18} by 9\%, and the zero-energy cross section has reduced by 16\%.
The scattering lengths for Ne, Ar, Kr, and Xe are 1.76~a.u., 1.98~a.u., 2.06~a.u., and 2.12~a.u., respectively. 
The reduction of the scattering length from its frozen-target value~\cite{Swann18} increases with the nuclear charge of the target, from 13\% for Ne to  41\% for Xe. Indeed, the zero-energy cross section for Xe is approximately one third of its frozen-target value. These results show that the van der Waals--type interaction due to simultaneous distortion of the Ps and the target plays a very significant role in Ps-atom scattering and must be accounted for in order to obtain accurate results.  Its effect is greater for heavier, more polarizable atoms. Among other predictions, our calculations are closest to the stochastic-variational results of Mitroy and coworkers~\cite{Mitroy01,Mitroy03}.

\begin{figure*}
\centering
\includegraphics{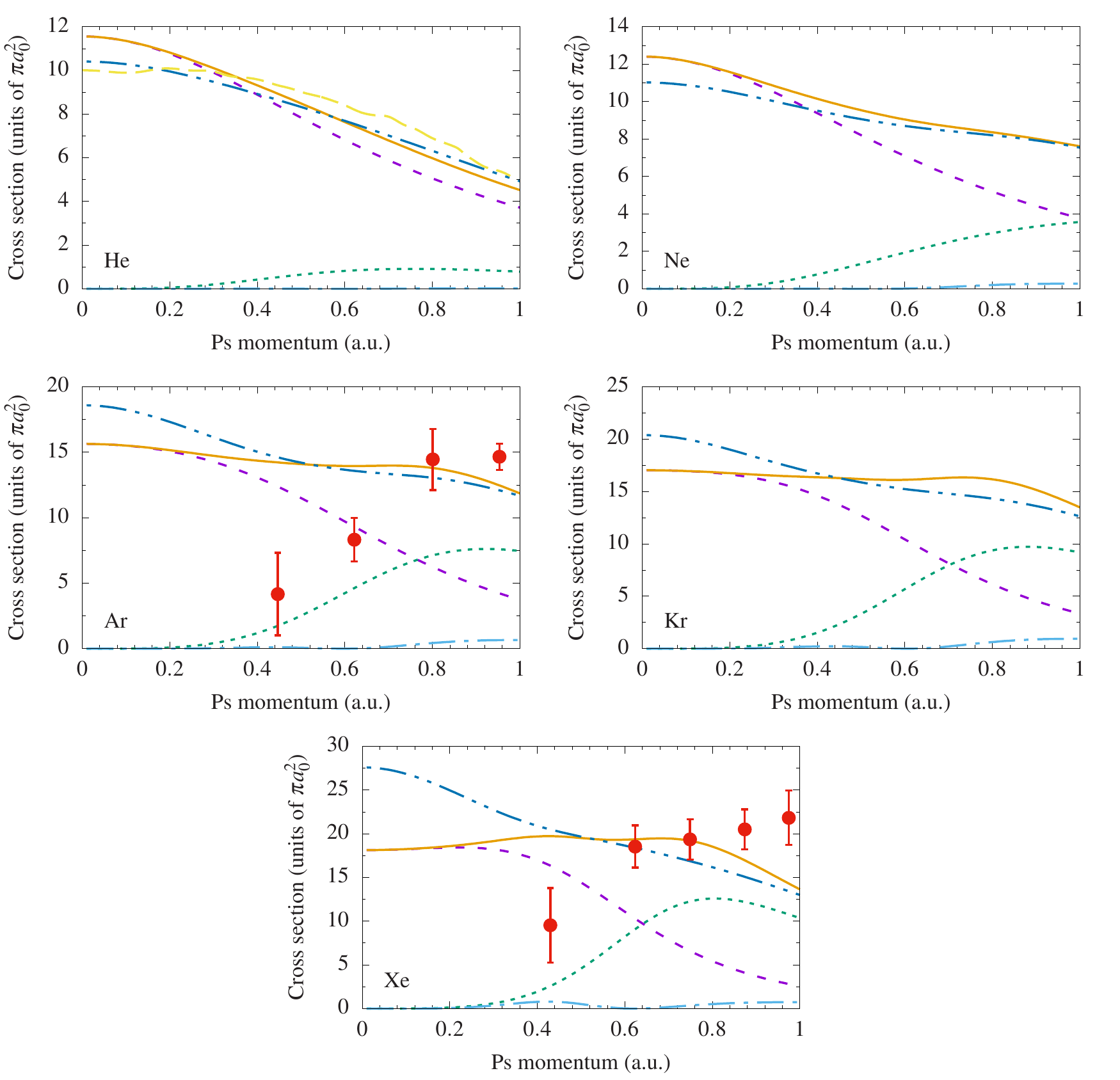}
\caption{\label{fig:XS_lp00-03hf}Elastic scattering cross sections for Ps scattering on He, Ne, Ar, Kr, and Xe. Present calculations: dashed purple lines, $S$-wave partial cross section; dotted green lines, $P$-wave partial cross section; dot-dashed light blue lines, $D$-wave partial cross section; solid orange lines, total cross section. Additional calculations of total cross section: dot-dash-dotted dark blue lines, van der Waals calculations with $R_0=3.0$~a.u.~\cite{Swann18}; long-dashed yellow line, 9-Ps--9-He coupled-channel calculation~\cite{Walters04}. The red circles are the experimental measurements of the total cross section by Brawley \textit{et al.}~\cite{Brawley15}.}
\end{figure*}

Figure~\ref{fig:XS_lp00-03hf} shows the elastic scattering cross sections as functions of the Ps center-of-mass momentum $K$, up to $K=1$~a.u. (the Ps breakup threshold).
They are generally rather flat and featureless curves, and they come into good agreement with the previous van der Waals calculations~\cite{Swann18} at high Ps momenta. The cross section for He is also in good agreement with the 9-Ps--9-He--state calculation of Walters \textit{et al.}~\cite{Walters04}, with only a 10\% discrepancy near $K=0$.

As we move along the noble gases in order of increasing nuclear charge, the behavior of the cross section at low energy goes from being decreasing for He, Ne, and Ar to almost flat for Kr, and to gently increasing for Xe. This is caused by the increasing contribution of $P$-wave scattering at intermediate momenta. Also, the valence electrons are less strongly bound for heavier target atoms (heavier target atoms have lower ionization potentials), so at low Ps energies, virtual excitations of Ps and the valence atomic electrons make for stronger attraction than for lighter target atoms.

The level of agreement with the available experimental data for Ar and Xe~\cite{Brawley15} is mixed. The contribution of the $D$ wave remains small for all atoms. For Xe there is close agreement for $0.6\lesssim K \lesssim 0.8$~a.u. However, the experimental datum for $K= 0.44$~a.u. is about a factor of 2 lower than our theoretical prediction, and the two experimental data with the highest momenta indicate an increase in the cross section, in contrast to our calculation. The $n=2$ excitation threshold for Ps lies at $K=\sqrt{3}/2\approx0.87$~a.u., so it is possible that for $K>0.87$~a.u. inelastic scattering with excitation of Ps($n=2$) states, which is neglected in our calculations, contributes to the experimental data.

Curiously, for Ar, Kr, and especially Xe, our cross sections begin to decrease markedly at $K\approx0.8$~a.u. Behavior of this type was not seen in either the frozen-target or van der Waals calculations (although the frozen-target and van der Waals cross sections were much more strongly decreasing across all momenta)~\cite{Swann18}. It appears to occur because the $P$-wave partial cross section reaches its maximum and begins to decrease at $K\approx0.8$--0.9~a.u., while the $D$-wave partial cross section remains almost insignificantly small---much smaller than in the frozen-target or van der Waals calculations~\cite{Swann18} (due to cancellation of the static repulsion and correlation attraction). It is, however, possible that our calculations become less accurate at momenta $K>0.8$~a.u., where the Ps energy is close to the Ps($n=2$) excitation threshold.

\begin{figure}
\centering
\includegraphics{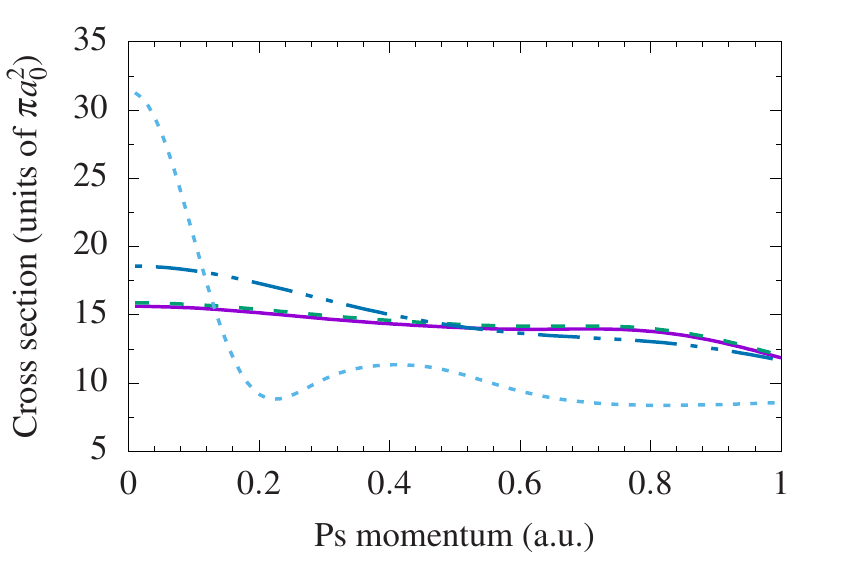}
\caption{\label{fig:Ar_different_approximations}Cross section for Ps scattering on Ar in various types of approximation. Solid purple line, inclusion of direct screening diagram only; dashed green line, inclusion of direct and exchange screening diagrams; dotted light blue line, inclusion of neither the direct nor exchange screening diagrams. The dot-dash-dotted dark blue line is the previous van der Waals calculation with $R_0=3.0$~a.u.~\cite{Swann18}.
}
\end{figure}

The effect of including or excluding the exchange screening diagrams (see Fig.~\ref{fig:screening_diagrams}) on the scattering cross section has been investigated for Ar.
Figure~\ref{fig:Ar_different_approximations} shows the total cross section for  Ar in three different approximations: including only the direct screening diagram, including the direct and exchange diagrams, and including neither the direct nor exchange screening diagrams (i.e., treating the interaction between the electron and positron as just the bare Coulomb interaction).
Also shown is the previous van der Waals calculation with $R_0=3.0$~a.u.~\cite{Swann18}.
It is evident that including only the direct screening diagram or including the direct and exchange screening diagrams gives very a similar cross section, differing by no more than 2\% across the momentum range considered.  It is therefore justified to exclude the exchange screening diagrams, on account of the computational expense required to calculate them. 
 Note that these  cross sections have the same basic shape as the earlier van der Waals cross section~\cite{Swann18}, coming into very close agreement for $K\gtrsim 0.5$~a.u. 
It is  quite remarkable that using a simple, local van der Waals potential with a single fitting parameter gives the cross section in such close agreement with the sophisticated \textit{ab initio} many-body-theory calculation. One may conclude from this that this potential (with a judiciously chosen cutoff radius $R_0$) captures well the physics of the Ps-atom interaction.
The calculation where neither the direct or exchange screening diagrams are included is in stark contrast to the others; the zero-energy cross section is more than a factor of 2 larger, and at $K\approx 0.2$~a.u. it is almost a factor of 2 lower. It also has a distinctly different momentum dependence. It is the least computationally expensive of the three types of approximation, but it is theoretically inconsistent, having neglected the effects of screening altogether. It serves as  a useful illustration of the importance of accounting for screening (which ensures the correct $-C_6/R^6$ asymptotic behavior of the Ps-atom interaction) in order to obtain reliable results. The unusual behavior is caused by the interplay of the $S$- and $P$-wave contributions to the total cross section. 
At low $K$, the $S$-wave phase shift $\delta_0$ is actually positive (the scattering length is negative, $A=-2.80$~a.u.). At $K\approx 0.28$~a.u., $\delta_0$ passes through zero and becomes negative, so the $S$ wave gives no contribution to the total cross section at this point (producing a Ramsauer-Townsend minimum). Coincidentally, the $P$-wave partial cross section has a maximum at almost the same momentum, offsetting the zero contribution of the $S$  wave. 

\begin{figure*}
\centering
\includegraphics{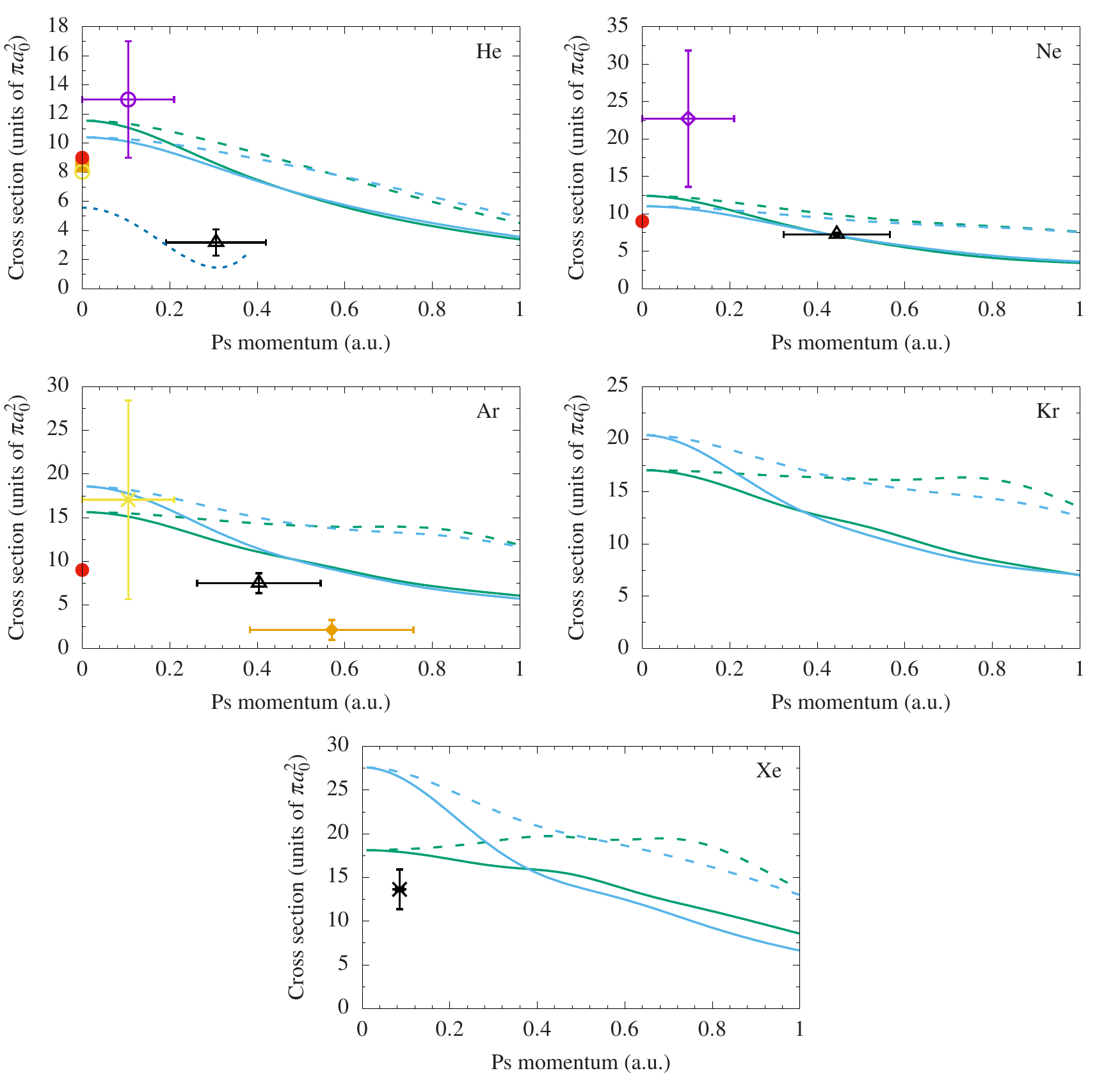}
\caption{\label{fig:MTXS_lp00-03hf}Momentum-transfer and total elastic cross sections for Ps scattering on He, Ne, Ar, Kr, and Xe. Solid (dashed) green lines, momentum-transfer (elastic) cross section using many-body theory; solid (dashed) light blue lines, momentum-transfer (elastic) cross section using model van der Waals potential with $R_0=3.0$~a.u. Experimental measurements of momentum-transfer cross section: filled orange square, Canter \textit{et al.}~\cite{Canter75}; open yellow circle, Rytsola \textit{et al.}~\cite{Rytsola84}; filled red circle, Coleman \textit{et al.}~\cite{Coleman94}; open black triangle, Skalsey \textit{et al.}~\cite{Skalsey03}; open purple circle, Nagashima \textit{et al.}~\cite{Nagashima98}; dotted dark blue line, Engbrecht \textit{et al.}~\cite{Engbrecht08}; open purple diamond, Saito \textit{et al.}~\cite{Saito03}; yellow cross, Nagashima \textit{et al.}~\cite{Nagashima95}; filled orange diamond, Sano \textit{et al.}~\cite{Sano15}; black asterisk, Shibuya \textit{et al.}~\cite{Shibuya13a}.}
\end{figure*}

Figure~\ref{fig:MTXS_lp00-03hf} shows the momentum-transfer cross sections for He, Ne, Ar, Kr, and Xe,
along with the corresponding van der Waals calculations using $R_0=3.0$~a.u.~\cite{Swann18}. For comparison, the elastic cross sections are also shown.
Moving away from $K=0$, the momentum-transfer cross sections all drop below the corresponding  elastic cross sections rather rapidly (due to destructive interference of the $S$- and $P$-wave contributions), and the many-body-theory and van der Waals calculations coalesce at high $K$. Also shown in  Fig.~\ref{fig:MTXS_lp00-03hf} are the experimental data from several groups, which are not direct measurements but the results extracted from observation of Ps thermalization in the noble gases.

 For He, our calculation lies within the error bars of the experimental result of Nagashima \textit{et al.}~\cite{Nagashima98} but is about 30--45\% higher than the earlier measurements of Canter \textit{et al.}~\cite{Canter75}, Rytsola \textit{et al.}~\cite{Rytsola84}, and Coleman \textit{et al.}~\cite{Coleman94}. The measurements of Skalsey \textit{et al.}~\cite{Skalsey03} and Engbrecht \textit{et al.}~\cite{Engbrecht08} give much lower values, and  the zero-energy cross section according to Engbrecht \textit{et al.}~\cite{Engbrecht08} is not consistent with the measurements of Canter \textit{et al.}~\cite{Canter75}, Rytsola \textit{et al.}~\cite{Rytsola84}, and Coleman \textit{et al.}~\cite{Coleman94}. As we noted in Ref.~\cite{Green18}, this may be because the meaurements of Skalsey \textit{et al.}~\cite{Skalsey03} and Engbrecht \textit{et al.}~\cite{Engbrecht08} are based on Doppler-broadening spectroscopy, and they may suffer from errors related to the discrimination of the narrow Ps annihilation component on the background of the positron-He annihilation signal.
 
For Ne, there is  agreement with the results of Skalsey \textit{et al.}~\cite{Skalsey03}, but our calculations lie outside the error bars of the measurement of Saito \textit{et al.}~\cite{Saito03}, and we have an approximately 40\% discrepancy with  Coleman \textit{et al.}~\cite{Coleman94}. 
For Ar, there is  agreement with the measurement by Nagashima \textit{et al.}~\cite{Nagashima95}, but the discrepancy with the results of Coleman \textit{et al.}~\cite{Coleman94}, Skalsey \textit{et al.}~\cite{Skalsey03}, and Sano \textit{et al.}~\cite{Sano15} is at the level of 70, 60, and 300\%, respectively. We again note that the general trend of the experimental data for Ar is a rapid decrease of the momentum-transfer cross section with increasing energy, which is qualitatively similar to our results but faster than the calculation predicts. 
Finally, for Xe, our momentum-transfer cross section is approximately 30\% larger than the experimental finding of Shibuya \textit{et al.}~\cite{Shibuya13a}. However, as we explained in Ref.~\cite{Swann18}, Shibuya \textit{et al.} deduced this datum by comparing the measured time evolution of the Ps center-of-mass energy $E(t)$ with the classical model described in Refs.~\cite{Sauder68,Nagashima95a}, viz.,
\begin{equation}
E(t) = E_\text{th} \coth^2(\alpha+\beta t),
\end{equation}
where $t$ is the time, $E_\text{th}=E(\infty)=\frac32 k_B T$ is the thermal energy at temperature $T$ (with $k_B$ the Boltzmann constant), $\alpha=\coth^{-1}\sqrt{E(0)/E_\text{th}}$, $\beta=p_\text{th} \sigma_m n / M$, $p_\text{th}$ is the thermalized Ps momentum, $n$ is the gas number density, and $M$ is the mass of the atom (Xe).
The values of $\alpha$ and $\beta$ were obtained from a least-squares fit of the measured $E(t)$ as $\alpha=0.094$ and $\beta=0.011$~ns$^{-1}$. However, their fit was based only on the restricted set of data ($40\leq E(t) \leq 60$~meV). Making use of all of the data points for $40\leq E(t) \leq 150$~meV, we found $\alpha=0.531$ and $\beta=0.0185$~ns$^{-1}$, giving a momentum-transfer cross section of $\sigma_m\approx 28\pi a_0^2$ at a Ps momentum of $K\sim 0.1$~a.u., which is about 50\% larger than our value.

\subsection{Pickoff annihilation}

Table~\ref{tab:oze_parameters} shows the values of $\oze(0)$ and $\oze'$ obtained [see Eq.~(\ref{eq:quadratic_oze})] and provides a   comparison with  existing calculations and experimental data. We carried out three sets of calculations for each target atom: frozen-target calculations, where we excluded $\Sigma_E$ for both the electron and positron and $\delta V_E$ from Eq.~(\ref{eq:bethe_salpeter}) and set $\gamma_{nl}=1$ for all atomic orbitals and positron angular momenta in Eq.~(\ref{eq:enhanced_oze}); \textit{unenhanced} many-body-theory calculations, where we included $\Sigma_E$ for both the electron and positron and $\delta V_E$ in Eq.~(\ref{eq:bethe_salpeter}) but kept $\gamma_{nl}=1$ for all atomic orbitals and positron angular momenta in Eq.~(\ref{eq:enhanced_oze}); and \textit{enhanced} many-body-theory calculations, where we included $\Sigma_E$ for both the electron and positron and $\delta V_E$ in Eq.~(\ref{eq:bethe_salpeter}) and use the values of $\gamma_{nl}$ given in Table~\ref{tab:enhancement_factors} in Eq.~(\ref{eq:enhanced_oze}).
\begin{table}
\centering
\caption{\label{tab:oze_parameters}Pickoff annihilation parameters $\oze(0)$ and $\oze' $ for Ps collisions with noble-gas atoms. The abbreviation VE means ``vertex enhancement''.}
\begin{ruledtabular}
\begin{tabular}{lcc}
 Method & $\oze(0)$ & $\oze' $ \\
\hline\\[-1.2ex]
\multicolumn{3}{c}{Ps-He collisions} \\
 Present, frozen target & 0.0273 & 0.0101 \\
 Present, many-body theory without VE & 0.0411 & 0.00281 \\
 Present, many-body theory with VE & 0.131 & 0.00809 \\
  Static exchange (not converged)~\cite{Fraser66} & 0.0177 \\
  Static exchange~\cite{Fraser68} & 0.033 \\
 Static exchange~\cite{Barker68} & 0.0347 \\
  Static exchange with van der Waals~\cite{Barker69} & 0.0445 \\
  Kohn variational, static exchange~\cite{Drachman70} & 0.042 \\
  $T$ matrix, model static exchange~\cite{Biswas00} & ${\approx}0.11$ & ${\approx}1.4$ \\
  Stochastic variational, frozen target~\cite{Mitroy01} & 0.0287 & 0.0044 \\
  Stochastic variational, van der Waals~\cite{Mitroy01} & 0.0378 & $-0.0152$ \\
  Experiment~\cite{Charlton85} & 0.125 \\[5pt]
\multicolumn{3}{c}{Ps-Ne collisions} \\
 Present, frozen target & 0.0512 & 0.0170 \\
  Present, many-body theory without VE & 0.0932 & $-0.00482$ \\
  Present, many-body theory with VE &  0.255 & $-0.0315$ \\
  Stochastic variational, frozen target~\cite{Mitroy01} & 0.0533 & 0.0100 \\
  Stochastic variational, van der Waals~\cite{Mitroy01} & 0.0922 & $-0.0717$ \\
  Experiment~\cite{Charlton85} & 0.235 \\[5pt]
\multicolumn{3}{c}{Ps-Ar collisions} \\
 Present, frozen target & 0.0316 & 0.0253 \\
  Present, many-body theory without VE & 0.103 & $-0.0836$ \\
  Present, many-body theory with VE &  0.516 & $-0.448$ \\
  Stochastic variational, frozen target~\cite{Mitroy01} & 0.0340 & 0.0084 \\
  Stochastic variational, van der Waals~\cite{Mitroy01} & 0.0964 & $-0.168$ \\
  Experiment~\cite{Charlton85} & 0.314 \\[5pt]
\multicolumn{3}{c}{Ps-Kr collisions} \\
 Present, frozen target & 0.0304 & 0.00687 \\
  Present, many-body theory without VE & 0.109 & $-0.111$ \\
  Present, many-body theory with VE &  0.678 & $-0.731$ \\
  Stochastic variational, frozen target~\cite{Mitroy03} & 0.0300 & 0.0247 \\
  Stochastic variational, van der Waals~\cite{Mitroy03} & 0.0913 & $-0.211$ \\
  Experiment~\cite{Saito06} & 0.36 \\[5pt]
\multicolumn{3}{c}{Ps-Xe collisions} \\
 Present, frozen target & 0.0261 & $0.00693$ \\
  Present, many-body theory without VE & 0.114 & $-0.151$ \\
 Present, many-body theory with VE &  0.939 & $-1.24$ \\
  Stochastic variational, frozen target~\cite{Mitroy03} & 0.0223 & 0.0165 \\
  Stochastic variational, van der Waals~\cite{Mitroy03} & 0.0891 & $-0.318$ \\
  Experiment~\cite{Saito06} & 0.48 \\
\end{tabular}
\end{ruledtabular}
\end{table}

For He, where several static-exchange calculations of $\oze(0)$ are available, the present frozen-target results differ from those of Refs.~\cite{Fraser68}, \cite{Barker68}, \cite{Drachman70}, and \cite{Biswas00} by 17\%, 21\%, 35\%, and 75\%, respectively. A difference of about $20$\% is not unexpected, because unlike the present frozen-target calculations, the static-exchange calculations do not account for distortion of the Ps projectile. As for the much larger discrepancies with Refs.~\cite{Drachman70,Biswas00}, it was pointed out earlier by Mitroy and Ivanov~\cite{Mitroy01}  that the model exchange interaction used in Ref.~\cite{Biswas00}  was of ``dubious validity,'' and the fact that the value of $\oze(0)$ in Ref.~\cite{Biswas00} is in excellent agreement with experiment~\cite{Charlton85} is  a coincidence. In fact, this static-exchange calculation is in poor agreement with the other static-exchange calculations~\cite{Fraser68,Barker68}), and one of the authors of Ref.~\cite{Drachman70} later admitted that an assumption made therein (that the direct potential is negligible compared with the exchange potential) was not quantitatively correct~\cite{DiRenzi03}.

Unlike the experimental data, which show an increase in $\oze(0)$ with the number of atomic electrons, our frozen-target value of $\oze(0)$ increases in moving from He to Ne and from Ne to Ar, but it decreases in moving from Ar to Kr and from Kr to Xe. The frozen-target calculations of Mitroy and coworkers~\cite{Mitroy01,Mitroy03} similarly increase from He through to Ar and then decrease from Ar through to Xe. These trends indicate that distortion of the target atom and short-range electron-positron correlations are much more important for heavier target atoms.

With regard to the experimental data~\cite{Charlton85,Saito06}, the frozen-target values of $\oze(0)$ are, unsurprisingly, an order of magnitude too small, which is expected from a calculation that neglects the distortion of the target atom and short-range electron-positron correlations.

Our many-body-theory calculations of $\oze(0)$, without account of annihilation vertex enhancement, are larger than the frozen-target values  by a factor that ranges from 1.5 (for He) to 4.4 (for Xe). These increased values of $\oze(0)$ are in agreement at the level of 1--28\% with the stochastic-variational calculations of Mitroy and coworkers~\cite{Mitroy01,Mitroy03} with inclusion of the van der Waals interaction. They still underestimate the experimental data~\cite{Charlton85,Saito06} by a factor of 2--4 as a result of the missing short-range electron-positron correlation effects.

Including the vertex enhancement factors in Eq.~(\ref{eq:enhanced_oze}) results in an order-of-magnitude increase in $\oze(0)$ from the frozen-target value  for all of the atoms. For He and Ne, there is excellent agreement with experiment~\cite{Charlton85}---at the level of 5\% and 9\%, respectively. In both cases, our calculation slightly overestimates the experiment. For Ar, Kr, and Xe, our calculations of $\oze(0)$ overestimate the experiment~\cite{Charlton85,Saito06} by factors of 1.6, 1.9, and 2.0, respectively. 
The value of $\oze'$ has changed sign from positive (in the frozen-target approximation) to negative for Ne, Ar, Kr, and Xe. 
Note that this sign change occurs even if the many-body-theory Ps wave function is used to calculate $\oze$ \textit{without} enhancement of the annihilation vertex (see Table~\ref{tab:oze_parameters}).
A possible explanation for this phenomenon is as follows. In the frozen-target approximation, the overall Ps-atom interaction is repulsive. With increasing energy, the Ps overcomes this repulsion to some extent and penetrates the target atom more, making pickoff annihilation more likely; hence $\oze'>0$. When the Dyson-orbital states and screening corrections are used to construct the Ps wave function, the dispersion interaction affects the Ps mostly at low energy; hence $\oze(0)$ increases significantly. At higher energy the effect of electron- and positron-atom correlations decreases, and values of $\oze$ get closer to what they were in the frozen-target approximation; hence $\oze'$ reduces (and happens to become negative for Ne, Ar, and Kr).

Our vertex-enhanced many-body-theory values of $\oze(0)$  increase monotonically with the number of target electrons, as is observed in experiment~\cite{Charlton85,Saito06}. However, our theoretical approach appears to  systematically overestimate the measured $\oze$, and the discrepancy increases with increasing number of atomic electrons. Possible explanations for this behavior are as follows:
\begin{enumerate}
\item The Ps  may be too strongly attracted to the target atom, i.e., the correlation potential for the electron and/or positron may be too strongly attractive. 
There is a possiblilty that the electron and positron self-energy and the screening correction should all be smaller due to the fact that the energy $E$ that appears in the denominators of the corresponding diagrams should be negative, rather than zero. 
This is investigated in Sec.~\ref{sec:results_energy_dependence}.
\item We may have overestimated the effect of short-range electron-positron correlations, i.e., the enhancement factors in Table~\ref{tab:enhancement_factors} could be too large. This could be partly due to using enhancement factors calculated using Hartree-Fock states rather than Dyson states. However, using the Dyson enhancement factors instead would still give values of $\oze$ significantly larger than experiment for Ar, Kr, and Xe. For example, for Xe, the enhancement factor that most significantly reduces in switching from Hartree-Fock to Dyson states is for an $s$-wave positron annihilating on the $5p$ orbital; its value changes from 9.26 to ${\approx}6$~\cite{Green18a}. Crudely scaling our current value of $\oze(0)$ by a factor of $6/9.26\approx0.65$ gives $\oze(0)\approx 0.61$, still 30\% larger than the experimental value of 0.48~\cite{Saito06}. And if we were to correctly change the  enhancement factors for each orbital and partial wave to their Dyson values separately, we would obtain a value even larger, as the enhancement factors for higher partial waves and/or core orbitals change less significantly.
\item The enhancement factors that were used were calculated for zero energy of the positron. However, when the positron is ``packaged'' within Ps, the positron states in the expansion (\ref{eq:Ps_wfn}) of the Ps wave function cover a  range of energies. It is possible that the contributions of higher-energy positron states should not be enhanced to the same degree as those of the lower-energy states. However, it is not clear why this effect would be more important for the heavier noble-gas atoms.
\end{enumerate}


\subsection{\label{sec:results_energy_dependence}Dependence of results on energy used in energy denominators of diagrams}

As we noted in Sec.~\ref{sec:en_dep_self_energy}, in calculating the self-energy of the electron or positron in the field of the atom and the screening corrections to the electron-positron Coulomb interaction within Ps, the energy $E$ appears in the energy denominators of the Goldstone diagrams, so we consistently calculated all of the self-energy and screening diagrams at $E=0$. We now briefly investigate how changing the value of $E$ affects the scattering cross section and value of $\oze(0)$ for each atom.

The true value of $E$ in the energy denominators should be $E=K^2/4-1/4-\Delta E$, where $K$ is the Ps center-of-mass momentum, $-\frac14$~a.u. is the internal energy of ground-state Ps, and $\Delta E$ estimates the typical excitation energy of the electron or positron within Ps. Thus, for low-energy collisions ($K\approx0$), the electron and positron self-energy diagrams and the screening corrections should be calculated for a negative energy of $-0.25$~a.u., or less. One can compute the mean value of $\Delta E$ for either the electron or positron in a particular Ps eigenstate in the cavity by
\begin{equation}
\langle \Delta E \rangle
=
\begin{cases} 
\sum_{\mu,\nu} \lvert C_{\mu\nu} \rvert^2 \epsilon_\mu & \text{(electron),} \\
\sum_{\mu,\nu} \lvert C_{\mu\nu} \rvert^2 \epsilon_\nu & \text{(positron),} 
\end{cases}
\end{equation}
where the $C_{\mu\nu}$ are the expansion coefficients in Eq.~(\ref{eq:Ps_wfn}), and $\epsilon_\mu$ ($\epsilon_\nu$) is the energy of electron (positron) state $\psi_\mu$ ($\psi_\nu$).
We calculated $\langle\Delta E\rangle$ for the electron and positron for the lowest-energy Ps eigenstate for He, Ar, and Xe (with $E=0$), and found typical values to be in the range 0.125--0.15~a.u. Indeed, for a sufficiently large cavity, the potential energy for the electron or positron is mostly due to the Coulomb interaction with the other particle (since in a large cavity, the Ps is statistically likely to be found far away from the target atom). The expected values of the  Ps kinetic energy $T$ and potential energy $V$ satisfy the virial theorem, $2\langle T\rangle=-\langle V\rangle$, so the total energy of the electron-positron pair is $\langle T\rangle+\langle V\rangle=\langle V\rangle/2$. This shows that for Ps moving with center-of-mass momentum $K\approx 0$, the expected value of the potential energy should be $-0.5$~a.u. (since the total Ps energy is just the internal energy, $-0.25$~a.u.) Finally, assuming that the electron and positron have approximately the same mean excitation energy $\langle\Delta E\rangle$, we find
$
2 \langle\Delta E\rangle - 0.5 \approx -0.25 ,
$
which gives $\langle\Delta E\rangle \approx 0.125$~a.u.

Taking into account the above considerations, we decided to  investigate the dependence of the scattering cross sections and value of $\oze$ in a pragmatic manner, by calculating the self-energy and screening diagrams at $E=-0.25$~a.u. and $E=-0.375$~a.u., in addition to the value of $E=0$ used previously.

\begin{figure*}
\centering
\includegraphics{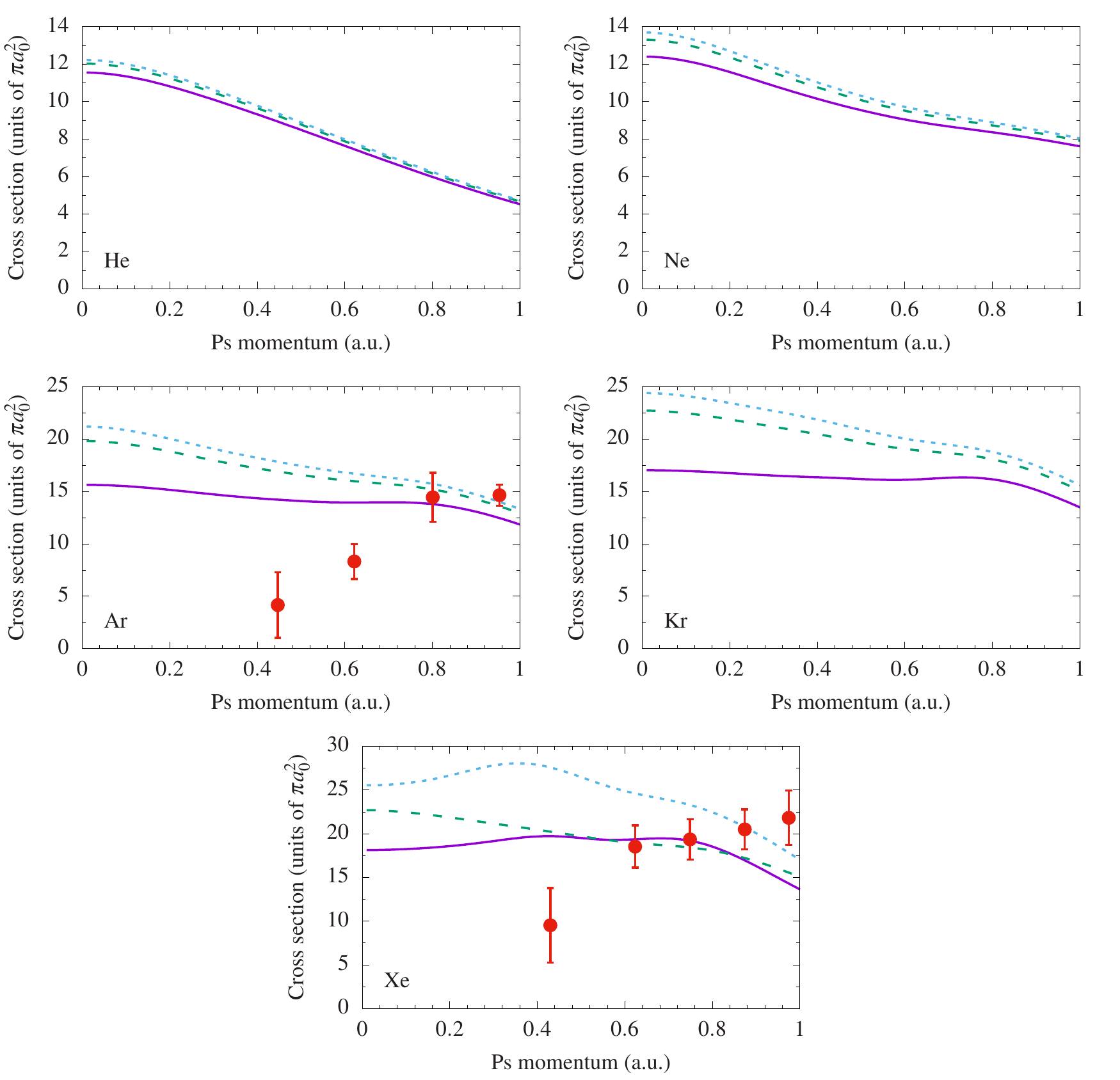}
\caption{\label{fig:XS_energy_dep}Elastic scattering cross sections for Ps scattering on He, Ne, Ar, Kr, and Xe, with the electron and positron self-energy diagrams and screening diagrams being calculated at various energies $E$. 
Solid purple lines, $E=0$; dashed green lines, $E=-0.25$~a.u.; dotted blue lines, $E=-0.375$~a.u.
The red circles are the experimental measurements  by Brawley \textit{et al.}~\cite{Brawley15}.}
\end{figure*}

Figure~\ref{fig:XS_energy_dep} shows the elastic cross sections for Ps scattering on He, Ne, Ar, Kr, and Xe for $E=0$, $-0.25$, and $-0.375$~a.u.
It is clear that for each atom, making the value of $E$ more negative results in a larger cross section (although for Xe, in the momentum range $K\approx 0.6$--0.85~a.u., the $E=-0.25$~a.u. cross section is smaller than the $E=0$ cross section, due to the delicate interplay of the $S$- and $P$-wave contributions). This occurs because making the energy denominators in the Goldstone diagrams more negative reduces their overall magnitude. Thus, the attraction of the electron and positron to the target atom (mainly due to the first, polarization-type diagrams in Fig.~\ref{fig:el_pos_diagrams}) is weakened, resulting in more negative phase shifts and a larger scattering length. The change in the cross section due to varying the value of $E$ is more pronounced at low Ps momenta $K$ and for heavier target atoms. For He, changing the value of $E$ from 0 to $-0.375$~a.u. increases the cross section at $K=0$ by 6\%, while the corresponding increase for Xe is 40\%. Despite the significant change for all atoms, the basic shape of the cross section remains the same for all values of $E$, and agreement with the experimental data~\cite{Brawley15} has not improved.

Table~\ref{tab:oze_ene_dep} shows the vertex-enhanced values of $\oze(0)$ for He, Ne, Ar, Kr, and Xe for for $E=0$, $-0.25$, and $-0.375$~a.u.
\begin{table}
\caption{\label{tab:oze_ene_dep}Values of $\oze(0)$, calculated using many-body theory with vertex enhancement, using various energies $E$ in the calculation of the self-energy and screening diagrams. Also shown are the experimental values.}
\begin{ruledtabular}
\begin{tabular}{lcccc}
& \multicolumn{3}{c}{$E$ (a.u.)} \\
\cline{2-4}
Atom & $0$ & $-0.25$ & $-0.375$ & Exp.~\cite{Charlton85,Saito06} \\
\hline
He & 0.131 & 0.127 & 0.124 & 0.125 \\
Ne & 0.255 & 0.239 & 0.231 & 0.235 \\
Ar & 0.516 & 0.409 & 0.375 & 0.314 \\
Kr & 0.678 & 0.497 & 0.446 & 0.36 \\
Xe & 0.939 & 0.607 & 0.530 & 0.48
\end{tabular}
\end{ruledtabular}
\end{table}
As expected, making $E$ more negative makes the value of $\oze(0)$ smaller, since the larger scattering length results in less penetration of Ps into the target atom. The values of $\oze(0)$ calculated for He and Ne with $E=-0.375$~a.u. are in near-perfect agreement with experiment~\cite{Charlton85}. For Ar, Kr, and Xe, the calculated values still overestimate the experimental data~\cite{Charlton85,Saito06} by factors of 1.2, 1.2, and 1.1, respectively, but these recommended values represent a very significant improvement on the $E=0$ values.

\section{Conclusions}

We have developed a many-body-theory approach to studying low-energy Ps interactions with noble-gas atoms. The entire Ps-atom system was enclosed in a hard-wall spherical cavity.
The Dyson equation was solved separately for the electron and positron moving in the field of the target atom, and the resulting states were used to construct the two-particle Ps wave function. 
Construction of the Ps wave function in this manner ensured that distortion of both the target atom and the Ps were accurately accounted for. The two-particle Dyson equation was solved, and the energies and wave functions of the Ps eigenstates in the cavity were used to determine the scattering phase shifts (and hence the cross sections) and pickoff annihilation parameter $\oze$.

Our calculations of the scattering cross section did not agree with the experimental data for Ar and Xe of Brawley \textit{et al.}~\cite{Brawley15}. In fact, it was found that the general shape of each cross section was similar to that predicted by simpler calculations that mimicked distortion of the target atom using a long-range model van der Waals potential~\cite{Swann18}, and the many-body-theory and van der Waals calculations generally came into good agreement for Ps momenta greater than 0.5~a.u. (at lower momenta, the van der Waals cross sections are quite sensitive to the choice of cutoff radius used in the model potential~\cite{Swann18}). Agreement with measurements of the momentum-transfer cross sections was also found to be mixed, though we note that measurements by different experimental groups are not always consistent.

Calculations of $\oze$ in the frozen-target approximation (performed by constructing the Ps wave function from Hartree-Fock electron and positron states in the field of the target atom) gave values an order of magnitude below the experimental value, due to the neglect of the effects of target-atom distortion and short-range electron-positron correlations. Accounting for target-atom distortion by constructing the Ps wave function from electron and positron Dyson states, and accounting for  the short-range electron positron correlations by scaling the atomic-orbital-- and positron-partial-wave--specific contributions to $\oze$ by the annihilation vertex enhancement factors from Ref.~\cite{Green18a}, we found values of $\oze$ in excellent agreement with experiment for He and Ne, and within a factor of 2 for Ar, Kr, and Xe.

Finally, we investigated the effect of changing the energy $E$ at which the electron and positron self-energy diagrams and screening corrections to the electron-positron Coulomb interaction were calculated. A value of $E=0$ had been used throughout, but the calculations were repeated for $E=-0.25$ and $-0.375$~a.u. Making $E$ more negative resulted in an overall less attractive Ps-atom interaction; thus, the scattering cross sections  increased and the values of $\oze$ decreased.  In going from $E=0$ to $E=-0.375$~a.u., the zero-energy cross section increased by 6--40\%, with the smallest (greatest) increase for He (Xe). However, the basic shape of the cross sections was unchanged, and no better agreement with the experimental data for Ar and Xe~\cite{Brawley15} was obtained. As for the pickoff annihilation parameter $\oze$, in going from $E=0$ to $E=-0.375$~a.u., the zero-energy values decreased by 5--45\%, with the smallest (greatest) decrease  for He (Xe). This brought the values of $\oze$ for He and Ne into near-exact agreement with experiment~\cite{Charlton85} and within a factor of 1.2 of experiment~\cite{Charlton85,Saito06} for Ar, Kr, and Xe.
Our calculations of $\oze$ for Ps collisions with noble-gas atoms are the only ones in the literature in which the short-range electron-positron correlation effects have been accounted for, and serve as current benchmarks. 

The work presented here represents a major step forward in our understanding of low-energy Ps-atom interactions: virtual target-atom excitations have now been included in Ps scattering by heavy atoms, and short-range electron-positron correlations have been accounted for in $\oze$. However, there are many ways in which the calculations could be improved and extended. In the self-energy expansion for the electron-atom interaction, only diagrams up to second order in the electron-electron Coulomb interaction were included; similarly, only second-order diagrams were included in the screening correction to the electron-positron Coulomb interaction within Ps. Though technically cumbersome, it is possible to implement higher-order corrections and check whether these significantly affect the Ps-atom scattering cross section and pickoff annihilation rate. Also, the many-body theory used in the calculations is nonrelativistic: all of the Goldstone diagrams were calculated using single-particle Hartree-Fock electron and positron orbitals. Although we expect relativistic corrections to be small for the Ps-atom problem, in principle, Dirac-Fock orbitals could be used to account for these effects. Further, we have only considered elastic scattering of Ps by noble-gas atoms. By investigating the nature of higher-energy states of Ps in the field of the target atom within the cavity, it may be possible to obtain information about inelastic scattering processes, e.g., excitation and/or ionization of the Ps and/or target atom. Lastly, although we have used the two-particle Ps wave function to calculate the pickoff annihilation parameter $\oze$ for each atom, it could also be used to compute the cross section of spin-orbit quenching, a spin-changing collision that transforms orthopositronium (with mean lifetime 142~ns) into parapositronium (with mean lifetime 0.125~ns), which rapidly annihilates into two or more $\gamma$ rays~\cite{Mitroy03b}.

The persisting discord between calculated low-energy Ps-atom scattering cross sections and experimental data is a cause for concern. From the point of view of theory, to produce a Ramsauer-Townsend minimum in the cross section, as predicted for Ar and Xe in Ref.~\cite{Brawley15}, would require the overall Ps-atom interaction at low energy to be attractive (indicated by a positive $S$-wave phase shift, giving a negative scattering length), becoming repulsive at intermediate energies (indicated by the $S$-wave phase shift becoming negative). However, the excellent agreement we obtained with experiment for $\oze$ leads us to believe that our many-body-theory approach broadly captures the essential physics of the Ps-atom system correctly. We hope that ongoing theoretical and experimental investigations will resolve the discrepancies for the cross sections in the near future.

\begin{acknowledgments}
The work of A.R.S. has been supported by the Department for the Economy, Northern Ireland, UK, and the EPSRC UK, Grant No. EP/R006431/1. The work of D.G.G. has been supported by the EPSRC UK, Grant No. EP/N007948/1 and ERC StG 804383. 
\end{acknowledgments}

\appendix

\begin{widetext}

\section{\label{sec:analytical_expressions}Analytical expressions for Goldstone diagrams and quantities derived from Ps wave function}

In calculating the various diagrams in $\langle \epsilon' \vert \Sigma_E \vert \epsilon \rangle$ and $\langle \epsilon' \vert \delta V_E \vert \epsilon \rangle$, integration over the angular variables is performed analytically~\cite{Varshalovich}. The Coulomb matrix element $\pm\langle \nu' \mu' \vert V \vert \mu \nu \rangle$ for two particles coupled to an angular momentum $J$ [see Eq.~(\ref{eq:coulomb_matrix_element})] is given by
\begin{equation}
\begin{gathered}
\includegraphics[scale=.7]{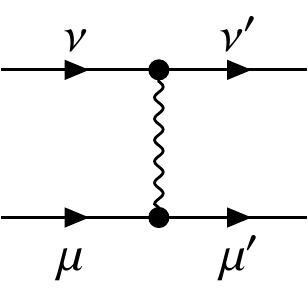} 
\end{gathered}
=
\pm\langle \nu' \mu' \vert V \vert \mu \nu \rangle
=
\pm
\sum_{l=0}^\infty (-1)^{J+l} 
\sj{J}{l_{\mu'}}{l_{\nu'}}{l}{l_\nu}{l_\mu}
\langle \nu' \mu' \Vert V_l \Vert \mu \nu \rangle ,
\end{equation}
where 
we choose the positive (negative) sign for a repulsive (attractive) Coulomb interaction,
the quantity in braces is a $6j$ symbol, the
 sum over $l$ is  actually finite (with the allowed values determined by the selection rules of the $6j$ symbol),  
\begin{equation}
\langle \nu' \mu' \Vert V_l \Vert \mu \nu \rangle
=
\sqrt{[l_{\nu'}][l_{\mu'}][l_\mu][l_\nu]}
\tj{l_{\nu'}}{l}{l_\nu}{0}{0}{0}
\tj{l_{\mu'}}{l}{l_\mu}{0}{0}{0}
\int_0^{R_c} \!\! \int_0^{R_c}
P_{\epsilon_{\nu'} l_{\nu'}}(r')
P_{\epsilon_{\mu'} l_{\mu'}}(r)
\frac{r_<^l}{r_>^{l+1}}
P_{\epsilon_{\mu} l_{\mu}}(r)
P_{\epsilon_{\nu} l_{\nu}}(r') \, dr \, dr'
\end{equation}
is the reduced Coulomb matrix element,  $[l]\equiv 2l+1$, $r_<=\min(r,r')$,  $r_>=\max(r,r')$, and the quantities in parentheses are $3j$ symbols. 

We now give expressions for the various Goldstone diagrams. Note that sums over magnetic quantum numbers and spins have already been carried out, so that sums over intermediate states only pertain to the radial and orbital quantum numbers, e.g., $\sum_{\mu} \equiv \sum_{\epsilon_\mu,l_\mu}$. For convenience, we define the following quantities which occur regularly:
\begin{align}
\langle \nu' \mu' \Vert V^{(l)} \Vert \mu \nu \rangle
&=
\sum_{l'=0}^\infty (-1)^{l+l'}
\sj{l}{l_{\mu'}}{l_{\nu'}}{l'}{l_\nu}{l_\mu}
\langle \nu' \mu' \Vert V_{l'} \Vert \mu \nu \rangle, \label{eq:red1}\\
\langle \nu' \mu' \Vert \widetilde{V}^{(l)} \Vert \mu \nu \rangle
&=
\sum_{l'=0}^\infty (-1)^{l+l'}
\sj{l}{l_{\mu'}}{l_\nu}{l'}{l_{\nu'}}{l_\mu}
\langle \nu' \mu' \Vert V_{l'} \Vert \mu \nu \rangle , \label{eq:red2}\\
\langle \nu'\mu' \Vert \Gamma_E^{(l)} \Vert \mu\nu \rangle
&=
-\langle \nu'\mu' \Vert V^{(l)} \Vert \mu\nu \rangle
-
\sum_{\mu'',\nu''}
\frac{\langle \nu'\mu' \Vert V^{(l)} \Vert \mu''\nu''\rangle \langle \nu'' \mu'' \Vert \Gamma_E^{(l)} \Vert \mu \nu \rangle}{E-\epsilon_{\mu''}-\epsilon_{\nu''}+i\delta} \label{eq:lad_expr} .
\end{align}
Note the different positions of $l_\nu$ and $l_{\nu'}$ in the $6j$ symbols in Eqs.~(\ref{eq:red1}) and (\ref{eq:red2}). Equation (\ref{eq:lad_expr}) is a linear matrix equation that can be solved to find the reduced ladder matrix elements $\langle \nu'\mu' \Vert \Gamma_E^{(l)} \Vert \mu\nu \rangle$. The expressions for the self-energy and screening diagrams are as follows:
\begin{align}
\begin{gathered}
\includegraphics[scale=.7]{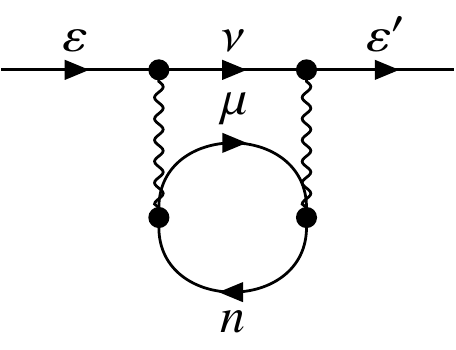} 
\end{gathered}
&=
\sum_{\substack{\mu,\nu>F \\ n\leq F}} \sum_l \frac{2}{[l][l_\epsilon]} \frac{\langle \epsilon' n \Vert V_l \Vert \mu\nu\rangle \langle  \nu \mu \Vert V_l \Vert n \epsilon \rangle}{E+\epsilon_n - \epsilon_\mu - \epsilon_\nu + i\delta} ,
\\
\begin{gathered}
\includegraphics[scale=.7]{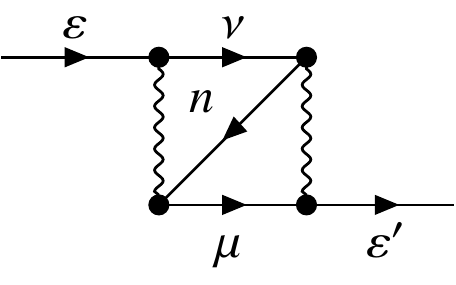} 
\end{gathered}
&=
-\sum_{\substack{\mu,\nu>F \\ n\leq F}} \sum_l
\frac{1}{[l_\epsilon]} \frac{\langle n \epsilon' \Vert \widetilde{V}^{(l)} \Vert \mu \nu \rangle \langle \nu \mu \Vert V_l \Vert n \epsilon \rangle}{E+\epsilon_n-\epsilon_\mu-\epsilon_\nu +i\delta} ,
\\
\begin{gathered}
\includegraphics[scale=.7]{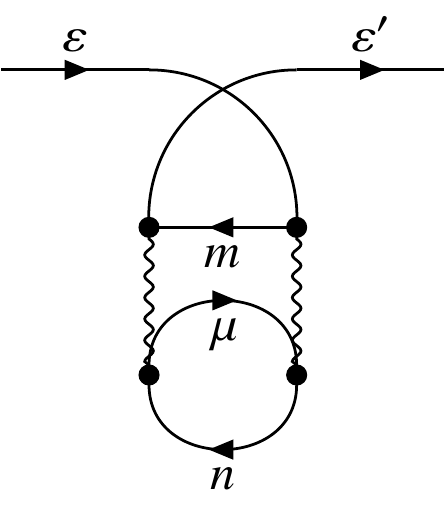} 
\end{gathered}
&=
- \sum_{\substack{\mu>F \\ m,n \leq F}} \sum_l
\frac{2}{[l][l_\epsilon]}
\frac{\langle \epsilon' \mu \Vert V_l \Vert n m \rangle \langle m n \Vert V_l \Vert \mu \epsilon \rangle}{-E+\epsilon_m+\epsilon_n-\epsilon_\mu+i\delta} ,
\\
\begin{gathered}
\includegraphics[scale=.7]{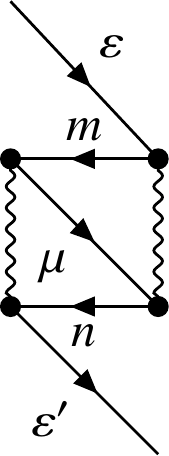} 
\end{gathered}
&=
\sum_{\substack{\mu>F \\ m,n \leq F}} \sum_l
\frac{1}{[l_\epsilon]}
\frac{\langle \mu \epsilon' \Vert \widetilde{V}^{(l)} \Vert n m \rangle \langle m n \Vert V_l \Vert \mu \epsilon \rangle}{-E+\epsilon_m+\epsilon_n-\epsilon_\mu+i\delta} ,
\\
\begin{gathered}
\includegraphics[scale=.7]{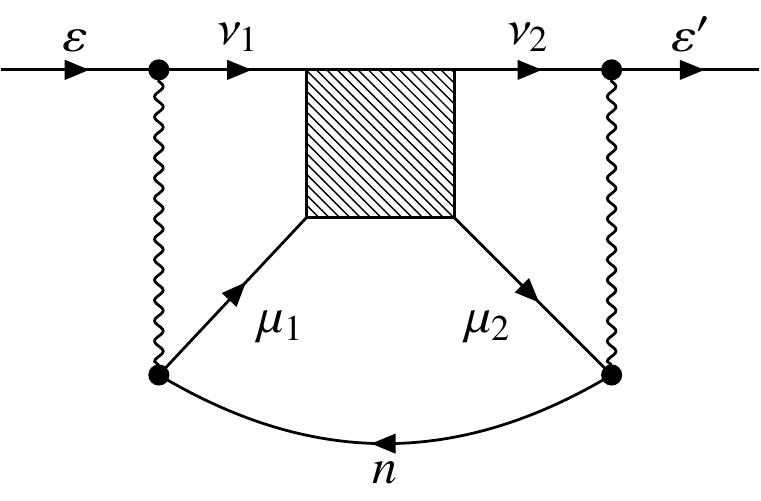} 
\end{gathered}
&=
\sum_{\substack{\mu_1,\mu_2>F \\ n\leq F}} \sum_{\nu_1,\nu_2} \sum_l
\frac{2 [l] }{[l_\epsilon]}
\frac{\langle \epsilon' n \Vert V^{(l)} \Vert \mu_2 \nu_2 \rangle \langle \nu_2 \mu_2 \Vert \Gamma_{E+\epsilon_n}^{(l)} \Vert \mu_1 \nu_1 \rangle \langle \nu_1 \mu_1 \Vert V^{(l)} \Vert n \epsilon \rangle}{(E+\epsilon_n-\epsilon_{\mu_1}-\epsilon_{\nu_1})(E+\epsilon_n-\epsilon_{\mu_2}-\epsilon_{\nu_2})} ,
\\
\begin{gathered}
\includegraphics[scale=.7]{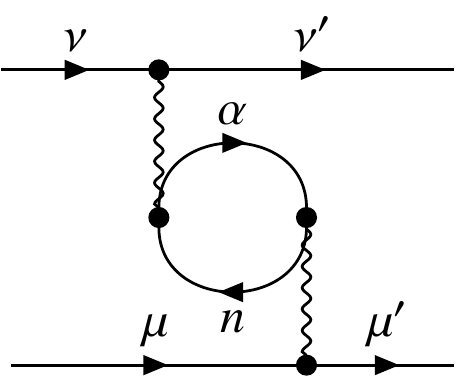} 
\end{gathered}
&=
-\sum_{\substack{\alpha>F \\ n\leq F}} \sum_l
\frac{1}{[l]}
\frac{\langle n \mu' \Vert V_l \Vert \mu\alpha \rangle \langle \alpha\nu'\Vert V_l \Vert \nu n \rangle}{E+\epsilon_n-\epsilon_\alpha+i\delta} ,
\\
\begin{gathered}
\includegraphics[scale=.7]{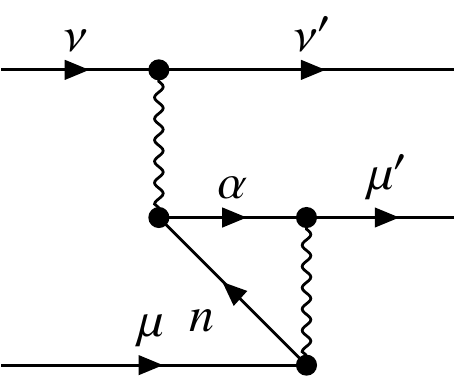} 
\end{gathered}
&=
 \sum_{\substack{\alpha>F \\ n\leq F}} \sum_l
\frac{\langle \mu' n \Vert V^{(l)} \Vert \mu \alpha \rangle \langle \alpha\nu'\Vert V_l \Vert \nu n\rangle}{E+\epsilon_n-\epsilon_\alpha+i\delta} ,
\\
\begin{gathered}
\includegraphics[scale=.7]{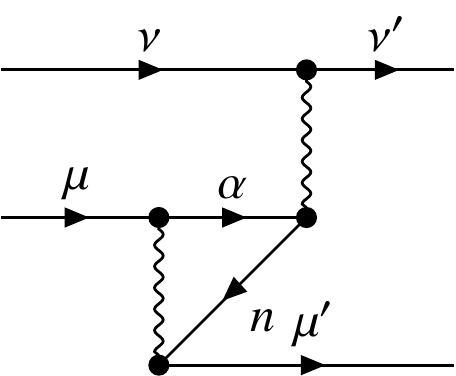} 
\end{gathered}
&=
 \sum_{\substack{\alpha>F \\ n\leq F}} \sum_l
\frac{\langle \mu'  \alpha  \Vert V^{(l)} \Vert \mu n \rangle \langle n \nu'\Vert V_l \Vert \nu \alpha\rangle}{E+\epsilon_n-\epsilon_\alpha+i\delta} ,
\\
\begin{gathered}
\includegraphics[scale=.7]{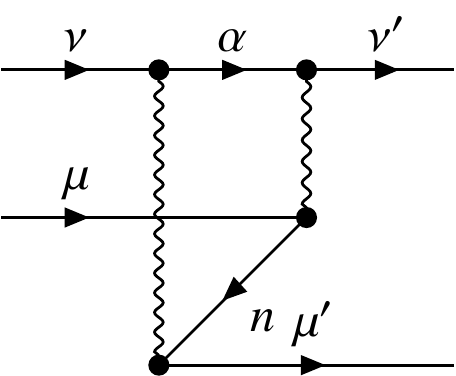} 
\end{gathered}
&=
-\sum_{n\leq F}\sum_\alpha\sum_{l,l',l''} 
(-1)^{l_\alpha+l_n} [l]
\sj{\mu'}{l}{\mu}{l'}{\alpha}{l''}
\sj{\nu'}{l}{\nu}{l''}{n}{l'}
\frac{\langle \nu' \alpha \Vert V_{l'} \Vert \mu n \rangle \langle n \mu' \Vert V_{l''} \Vert \alpha \nu \rangle}{E+\epsilon_n-\epsilon_\alpha+i\delta} .
\end{align}

The two-particle Ps wave function (\ref{eq:Ps_wfn}) is written explicitly in terms of the single-particle electron and positron basis states as
\begin{equation}\label{eq:Ps_wfn_explicit}
\Psi_{J\Pi}(\vec{r}_e,\vec{r}_p) = \frac{1}{r_e r_p} \sum_{\substack{\epsilon_\mu ,l_\mu \\ \epsilon_\nu , l_\nu}} C_{\epsilon_\mu l_\mu  \epsilon_\nu  l_\nu}^{(J\Pi)} \mathcal{P}_{\epsilon_\mu l_\mu}(r_e) \mathcal{P}_{\epsilon_\nu l_\nu}(r_p) \sum_{m_\mu,m_\nu} C_{l_\mu m_\mu l_\nu m_\nu}^{JM} Y_{l_\mu m_\mu}(\vec{\hat{r}}_e) Y_{l_\nu m_\nu}(\vec{\hat{r}}_p),
\end{equation}
where $C_{\epsilon_\mu l_\mu  \epsilon_\nu  l_\nu}^{(J\Pi)}$ is an expansion coefficient and $C_{l_\mu m_\mu l_\nu m_\nu}^{JM}$ is a Clebsch-Gordan coefficient \footnote{Comparing Eq.~(\ref{eq:Ps_wfn_explicit}) with Eq.~(\ref{eq:Ps_wfn}), $C_{\mu\nu}\equiv C_{\epsilon_\mu l_\mu  \epsilon_\nu  l_\nu}^{(J\Pi)}  \sum_{m_\mu,m_\nu} C_{l_\mu m_\mu l_\nu m_\nu}^{JM} $.}. Besides the selection rules due to the Clebsch-Gordan coefficient, the summation is restricted by parity, $(-1)^{l_\mu+l_\nu}=\Pi$, where $\Pi=1$ ($-1$) for the even (odd) states.

We now consider the calculation of $\oze$. The density of the atomic electrons  is
\begin{align}\label{eq:rhor}
\rho(\vec{r}) &= \sum_{n\leq F} \lvert \psi_n(\vec{r}) \rvert^2 
= 2\sum_{\epsilon_n ,l_n} \frac{[l_n]}{4\pi} \frac{\mathcal{P}_{\epsilon_nl_n}(r)^2}{r^2},
\end{align}
where the factor of 2 accounts for summation over the spins. Assuming that the electron and positron in Ps are coupled to an angular momentum of $J=0$, substituting Eq.~(\ref{eq:rhor}) into Eq.~(\ref{eq:enhanced_oze}), and carrying out the angular integrals  analytically, we obtain
\begin{align}
\oze = \sum_{\epsilon_n,l_n}  \frac{[l_n]}{8\pi} \sum_{\epsilon_\mu,\epsilon_{\nu},\epsilon_{\nu'},l} \gamma_{nl}
C_{\epsilon_\mu l \epsilon_\nu l}^{(0+)} C_{\epsilon_\mu l \epsilon_{\nu'} l}^{(0+)}
\int_0^{R_c} \mathcal{P}_{\epsilon_n l_n}(r)^2 \mathcal{P}_{\epsilon_\nu l}(r) \mathcal{P}_{\epsilon_{\nu'} l}(r) \frac{dr}{r^2}.
\end{align}

The method of calculating the Ps center-of-mass density $\rho_\text{cm}(\vec{r})$, Eq.~(\ref{eq:rhocm}), depends on whether $\vec{r}=\vec{0}$ or $\vec{r}\neq\vec{0}$. 
For $\vec{r}=\vec{0}$, the delta function can be expanded as
\begin{align}
\delta\left( \frac{\vec{r}_e+\vec{r}_p}{2}-\vec{0}\right) = 8 \frac{\delta(r_e-r_p)}{r_e^2} \sum_{l=0}^\infty (-1)^l \sum_{m=-l}^l Y_{lm}^*(\vec{\hat{r}}_e) Y_{lm} (\vec{\hat{r}}_p),
\end{align}
giving
\begin{align}
\rho_\text{cm}(\vec{0}) &= 
\sum_{\substack{\epsilon_\mu,l_\mu \\ \epsilon_\nu,l_\nu}}
\sum_{\substack{\epsilon_{\mu'},l_{\mu'} \\ \epsilon_{\nu'},l_{\nu'}}}
C^{(J\Pi)}_{\epsilon_\mu l_\mu \epsilon_\nu l_\nu}
C^{(J\Pi)}_{\epsilon_{\mu'} l_{\mu'} \epsilon_{\nu'} l_{\nu'}}
\sum_l
(-1)^{J}
\sj{J}{l_{\mu'}}{l_{\nu'}}{l}{l_\nu}{l_\mu}
\frac{2[l]}{\pi}
\sqrt{[l_{\nu'}][l_{\mu'}][l_\mu][l_\nu]}
\tj{l_{\nu'}}{l}{l_\nu}{0}{0}{0} 
\tj{l_{\mu'}}{l}{l_\mu}{0}{0}{0} 
\nonumber\\ &\quad{}\times
\int_0^{R_c}  \mathcal{P}_{\epsilon_{\nu'} l_{\nu'}}(r) \mathcal{P}_{\epsilon_{\mu'} l_{\mu'}}(r) 
 \mathcal{P}_{\epsilon_\mu l_\mu}(r) \mathcal{P}_{\epsilon_\nu l_\nu}(r) \frac{dr}{r^2}.
\end{align}
For $\vec{r}\neq\vec{0}$, the delta function expands as
\begin{align}\label{eq:A20}
\delta \left( \frac{\vec{r}_e+\vec{r}_p}{2} - \vec{r} \right)
= \frac{\delta(\lvert\vec{r}_e+\vec{r}_p)\rvert/2-r)}{r^2}
\sum_{l=0}^\infty \frac{[l]}{4\pi}P_l(\cos\omega),
\end{align}
where $P_l$ is the Legendre polynomial and $\omega$ is the angle between $\vec{r}_e+\vec{r}_p$ and $\vec{r}$. A difficulty arises in that the delta function on the RHS of Eq.~(\ref{eq:A20}) also needs to be expanded. For simplicity, we only consider the case where $J=0$. Then only the $l=0$ term on the RHS of Eq.~(\ref{eq:A20}) is nonzero:
\begin{align}\label{eq:A21}
\delta \left( \frac{\vec{r}_e+\vec{r}_p}{2} - \vec{r} \right)
= \frac{\delta(\lvert\vec{r}_e+\vec{r}_p\rvert/2-r)}{r^2}
\frac{1}{4\pi} = \frac{1}{2\pi r^2} \delta\left( \left\lvert \vec{r}_e+\vec{r}_p \right\rvert - 2r \right).
\end{align}
We now expand $\delta( \lvert \vec{r}_e+\vec{r}_p \rvert - 2r )$ as
\begin{align}\label{eq:A22}
\delta( \lvert \vec{r}_e+\vec{r}_p \rvert - 2r ) = \sum_{l=0}^\infty \frac{[l]}{4\pi} g_{l}(r_e,r_p) P_{l}(\cos\omega),
\end{align}
where the expansion coefficients $g_{l}$ are to be determined. Multiplying both sides of Eq.~(\ref{eq:A22}) by $P_{l'}(\cos\omega)\sin\omega$, integrating over $\omega$ from 0 to $\pi$, changing variables to $x\equiv\cos\omega$, and relabelling $l'$ as $l$, we obtain
\begin{align}
g_l(r_e,r_p) = 2\pi\int_{-1}^1 \delta \left( \sqrt{r_e^2+r_p^2+2r_er_px}-2r \right)P_l(x) \, dx.
\end{align}
We recall that a general property of the delta function is
\begin{align}
\delta[f(x)] = \sum_i \frac{\delta(x-x_i)}{\lvert f'(x_i) \rvert} ,
\end{align}
where the $x_i$ are the roots of $f(x)$. In this case we obtain
\begin{align}
g_l(r_e,r_p) = \frac{4\pi r}{r_e r_p} \int_{-1}^1 \delta \left( x-\frac{4r^2-r_e^2-r_p^2}{2r_er_p} \right) P_l(x) \, dx,
\end{align}
which gives
\begin{align}\label{eq:A26}
g_l(r_e,r_p) =
\frac{4\pi r}{r_er_p} P_l\left(\frac{4r^2-r_e^2-r_p^2}{2r_er_p} \right)
\end{align}
if $\lvert r_e-r_p\rvert < 2r < r_e+r_p$, and $g_l(r_e,r_p)=0$ otherwise.
Combining Eqs.~(\ref{eq:rhocm}), (\ref{eq:A21}), (\ref{eq:A22}), and (\ref{eq:A26}), and noting that for $J=0$ we require $l_\mu=l_\nu$ and $l_{\mu'}=l_{\nu'}$, we obtain
\begin{align}
\rho_\text{cm}(\vec{r}) &= 
\sum_{\substack{\epsilon_\mu,\epsilon_\nu \\ \epsilon_{\mu'},\epsilon_{\nu'} }}
\sum_{l_\mu,l_{\mu'}}
C^{(0+)}_{\epsilon_\mu l_\mu \epsilon_\nu l_\mu}
C^{(0+)}_{\epsilon_{\mu'} l_{\mu'} \epsilon_{\nu'} l_{\mu'}}
 (-1)^{l_\mu+l_{\mu'}} \sqrt{[l_\mu][l_{\mu'}]}
 \sum_l
[l]
\tj{l_{\mu'}}{l}{l_\mu}{0}{0}{0}^2
\nonumber\\ &\quad{}\times
\frac{1}{2\pi r}\int_0^{R_c} \!\! \int_{\vert 2r-r_p \vert}^{\min(2r+r_p,R_c)} 
\mathcal{P}_{\epsilon_{\nu'} l_{\mu'}}(r_p) 
\mathcal{P}_{\epsilon_{\mu'} l_{\mu'}}(r_e) 
P_l\left(\frac{4r^2-r_e^2-r_p^2}{2r_er_p} \right)
\mathcal{P}_{\epsilon_\mu l_\mu}(r_e) 
\mathcal{P}_{\epsilon_\nu l_\mu}(r_p)  \frac{dr_e}{r_e} \frac{dr_p}{r_p} .
\end{align}

\end{widetext}

\section{\label{sec:ert_fits}Effective-range-theory fits for scattering phase shifts}

The Ps scattering phase shifts are determined from the Ps energy
eigenvalues using the boundary condition on the Ps center-of-mass
motion at the cavity wall, as described in Ref.~\cite{Swann18}.
Calculations were performed using cavity radii of 10, 12, 14,
and 16 a.u. Effective-range-type fits were used to interpolate
the $S$, $P$, and $D$ phase shifts calculated at the discrete values
of the Ps center-of-mass momentum $K$. These fits were used
to determine the scattering length and the partial contributions
to the elastic and momentum-transfer cross sections. Here we detail the fits for the phase shifts.

At low momenta $K$, the $S$-wave phase shift $\delta_0$ behaves according to
\begin{equation}
K\cot\delta_0 \simeq -\frac1A + \frac12r_0K^2,
\end{equation}
where $A$ is the scattering length and $r_0$ is the effective range~\cite{Burke77}. This can be rearranged to give
\begin{equation}
    \delta_0(K) = \tan^{-1} \frac{K}{a_0 + a_1 K^2} \quad (\operatorname{mod}\pi) ,
\end{equation}
where $a_0=-1/A$ and $a_1=r_0/2$.
The $P$- and $D$-wave phase shifts behave according to \begin{align*}
    \delta_1 &= \alpha K^3 + \beta K^4 + \gamma K^5 + \epsilon K^7 \ln K + O(K^7), \\
    \delta_2 &= \zeta K^4 + \eta K^5 + \lambda K^7 + \mu K^9 \ln K + O(K^9),
\end{align*}
respectively, where $\alpha$, $\beta$, $\gamma$, $\epsilon$, $\zeta$, $\eta$, $\lambda$, and $\mu$ are constants~\cite{Burke77}. We found that simple polynomial fits obtained by truncating these expansions tend to grow large as $K\to1$~a.u., so we used the following Pad\'e-type fits instead:
\begin{align*}
\delta_1(K) &= \frac{a_0 K^3}{1 + a_1K^2 + a_2K^4} , \\
\delta_2(K) &= \frac{a_0K^4+a_1K^5}{1+a_2K^6}.
\end{align*}
These fits have the correct leading-order behavior as $K\to0$ and vary relatively slowly at large $K$, as observed in the calculated phase shifts.
Table~\ref{tab:phase_shifts}  shows the values of the parameters of the fits for $\delta_L$ ($L=0$, 1, 2) for He, Ne, Ar, Kr, and Xe.

\begin{table*}
\caption{\label{tab:phase_shifts}Parameters for the effective-range fits for the $S$-, $P$-, and $D$-wave phase shifts $\delta_L$.}
\begin{ruledtabular}
\begin{tabular}{ccd{2.3}d{-1.3}d{2.3}d{2.3}d{-2.3}}
&& \multicolumn{5}{c}{Atom} \\
\cline{3-7}
$L$ & Parameter & \multicolumn{1}{c}{\text{He}} & \multicolumn{1}{c}{\text{Ne}} & \multicolumn{1}{c}{\text{Ar}} & \multicolumn{1}{c}{\text{Kr}} & \multicolumn{1}{c}{\text{Xe}} \\
\hline
$0$ & $a_0$ & -0.588 & -0.568 & -0.506 & -0.485 & -0.470 \\
       & $a_1$ & 0.312 & 0.323 & 0.772 & 0.921 & 1.22 \\
$1$ & $a_0$ & -1.90 & -2.31 & -2.16 & -2.35 & -2.38 \\
       & $a_1$ & 3.41 & 2.86 & 0.439 & 0.167 & -0.607 \\
       & $a_2$ & 2.92 & 0.138 & 0.940 & 1.04 & 1.61 \\
$2$ & $a_0$ & 1.01 & 1.18 & 3.46 & 5.15 & 9.65 \\
       & $a_1$ & -1.73 & -2.25 & -5.78 & -8.52 & -15.5 \\
       & $a_2$ & 22.4 & 8.19 & 11.7 & 14.4 & 29.8 \\
\end{tabular}
\end{ruledtabular}
\end{table*}

\bibliography{Ps_MBPT_PRA_bib}

\end{document}